\documentstyle[12pt]{article}  
\begin{document}
\def\baselinestretch{1.2}
\def\thefootnote{\fnsymbol{footnote}}
\def\Tr{{\rm Tr}}
\def\[{\left [}
\def\]{\right ]}
\def\({\left (}
\def\){\right )}
\def\lbr{\left\{}
\def\rbr{\right\}}
\def\pp{\partial}
\def\hB{\hat{B}}
\def\D{{\cal D}}
\def\G{{\cal G}}
\def\tB{{\tilde B}}
\def\tb{{\tilde b}}
\def\tG{{\tilde\Gamma}}
\def\tP{{\tilde\Phi}}
\def\m{\bar{m}}
\newcommand{\be}{\begin{equation}}
\newcommand{\ee}{\end{equation}}
\newcommand{\bea}{\begin{eqnarray}}
\newcommand{\eea}{\end{eqnarray}}
\newcommand{\Lag}{{\cal L}}
\newcommand{\superint}{\int \diff^{4}\theta}
\newcommand{\lowest}{|_{\theta =\bar{\theta}=0}}
\newcommand{\diff}{\mbox{d}}
\newcommand{\Diff}{{\cal D}}
\newcommand{\WaWa}{\Tr({\cal W}^{\alpha}{\cal W}_{\alpha})}
\newcommand{\WbWb}{\Tr({\cal W}_{\dot{\alpha}}{\cal W}^{\dot{\alpha}})}
\newcommand{\DaDa}{{\cal D}^{\alpha}{\cal D}_{\alpha}}
\newcommand{\DbDb}{{\cal D}_{\dot{\alpha}}{\cal D}^{\dot{\alpha}}}
\newcommand{\Da}{{\cal D}^{\alpha}}
\newcommand{\Db}{{\cal D}_{\dot{\alpha}}}
\newcommand{\hs}{\hspace{0.2cm}}
\newcommand{\dg}{g_{_{(1)}}}
\newcommand{\dgg}{g_{_{(2)}}}
\newcommand{\dgm}{g_{_{(m)}}}
\newcommand{\df}{f_{_{(1)}}}
\newcommand{\dff}{f_{_{(2)}}}
\newcommand{\dfm}{f_{_{(m)}}}
\newcommand{\dilaton}{{\ell}}

\def\ksection{\arabic{section}}
\def\theequation{\ksection.\arabic{equation}}

 
\catcode`\@=11
 
\def\thesection{\arabic{section}.}
\def\thesubsection{\Alph{subsection}.}
\def\thesubsubsection{\roman{subsubsection}.}

\begin{titlepage}
\begin{center}
            \hfill    LBNL-39608 \\
            \hfill    UCB-PTH-96/54 \\
            \hfill hep-th/9611149\\

{\large \bf Modular Invariant Formulation of \\ Multi-Gaugino and Matter 
Condensation}\footnote{This work was supported in part by the 
Director, Office of 
Energy Research, Office of High Energy and Nuclear Physics, Division 
of High Energy Physics of the U.S. Department of Energy under 
Contract DE-AC03-76SF00098 and in part by the National Science 
Foundation under grant PHY-95-14797.}\\[.1in]

    Pierre Bin\'{e}truy,\footnote{Visiting Miller Professor. Permanent address:
Laboratoire de Physique Th\'{e}orique et Hautes Energies (Laboratoire 
associ\'e au CNRS--URA-D0063), 
Universit\'{e} Paris-Sud, F-91405 Orsay, France}
Mary K. Gaillard\footnote{Miller Professor, Fall 1996.} and Yi-Yen Wu 

{\em  Department of Physics,University of California, and 

 Theoretical Physics Group, 50A-5101, Lawrence Berkeley National Laboratory, 
      Berkeley, CA 94720, USA}
\end{center}

\begin{abstract}
      Using the linear multiplet formulation for the dilaton superfield,
we construct an effective lagrangian for hidden-sector gaugino condensation 
in string effective field theories with arbitrary gauge groups and matter. 
Nonperturbative string corrections to the K\"ahler potential are invoked to
stabilize the dilaton at a supersymmetry breaking minimum of the potential. 
When the cosmological constant is tuned to zero the
moduli are stabilized at their self-dual points, and the $vev$'s of their
F-component superpartners vanish.  Numerical analyses of 
one- and two-condensate examples with massless chiral matter
show considerable enhancement of the gauge hierarchy with respect to the $E_8$ 
case.  The nonperturbative string effects required for dilaton stabilization 
may have implications for gauge coupling 
unification.  As a comparison, we also consider a parallel approach based on
the commonly used chiral formulation.
 
\end{abstract}
\end{titlepage}
\renewcommand{\thepage}{\roman{page}}
\setcounter{page}{2}
\mbox{ }

\vskip 1in

\begin{center}
{\bf Disclaimer}
\end{center}

\vskip .2in

\begin{scriptsize}
\begin{quotation}
This document was prepared as an account of work sponsored by the United
States Government.  Neither the United States Government nor any agency
thereof, nor The Regents of the University of California, nor any of their
employees, makes any warranty, express or implied, or assumes any legal
liability or responsibility for the accuracy, completeness, or usefulness
of any information, apparatus, product, or process disclosed, or represents
that its use would not infringe privately owned rights.  Reference herein
to any specific commercial products process, or service by its trade name,
trademark, manufacturer, or otherwise, does not necessarily constitute or
imply its endorsement, recommendation, or favoring by the United States
Government or any agency thereof, or The Regents of the University of
California.  The views and opinions of authors expressed herein do not
necessarily state or reflect those of the United States Government or any
agency thereof of The Regents of the University of California and shall
not be used for advertising or product endorsement purposes.
\end{quotation}
\end{scriptsize}

\vskip 2in

\begin{center}
\begin{small}
{\it Lawrence Berkeley Laboratory is an equal opportunity employer.}
\end{small}
\end{center}

\newpage
\renewcommand{\thepage}{\arabic{page}}
\renewcommand{\theequation}{\arabic{section}.\arabic{equation}}
\setcounter{page}{1}
\def\thefootnote{\arabic{footnote}}
\setcounter{footnote}{0}

\section{Introduction}
Effective Lagrangians for gaugino condensation in effective field theories from
superstrings were first constructed by generalizing the work of Veneziano and 
Yankielowicz~\cite{vy}
to include the dilaton~\cite{tom} and gravity~\cite{bg89}. These constructions
used the chiral formulation for the dilaton superfield.  While the resulting 
Lagrangian has a simple interpretation~\cite{bg91} in terms of the two-loop
running of the gauge coupling constant, it does not respect the modular
invariance~\cite{mod} of the underlying superstring theory.  Modular invariance
was recovered~\cite{fmtv} by adding a moduli-dependent term to the
superpotential that is reminiscent of threshold corrections~\cite{dkl} found 
in some orbifold compactifications.  
However there is a large class of orbifolds that do
not have moduli-dependent threshold corrections~\cite{ant}; moreover in all
orbifold models, at least part of the modular anomaly is cancelled by a
Green-Schwarz counterterm~\cite{gs}, which must therefore be included.  This has
the unfortunate effect of destabilizing the dilaton.

It was recently shown how to formulate gaugino condensation using 
the linear multiplet~\cite{linear} formulation for the dilaton superfield, 
both in global
supersymmetry~\cite{bdqq,bgt} and in the superconformal formulation of
supergravity~\cite{bgt}.  In this case the superfield $U$ which is the
interpolating field for the Yang-Mills composite superfield ($U\simeq\WaWa$)
emerges as the chiral projection of a real vector supermultiplet 
$V$ whose lowest component is the dilaton field $\ell$.  
Using the K\"ahler superspace formalism of 
supergravity~\cite{bggm,bggml}, which we use throughout this paper, it was 
subsequently shown~\cite{bg96} how to include the Green-Schwarz term for a pure 
Yang-Mills
$E_8$ hidden sector.  In this case there are no moduli-dependent threshold
corrections and there is a single constant--the $E_8$ Casimir $C$--that governs
both the Green-Schwarz term and the coupling renormalization.
That model was studied in detail in~\cite{us}, where it was found that the
dilaton can be stabilized at a phenomenologically acceptable value with broken
supersymmetry if
nonperturbative terms~\cite{bd,shenk} are included in the K\"ahler 
potential,\footnote{A similar observation has been made by Casas~\cite{casas}
in the context of the chiral formulation and without modular invariance.}
but a sufficiently large gauge hierarchy is not generated.

The advantage of the linear multiplet formulation of gaugino condensation is
twofold. First, it is the correct string formulation since among the massless
string modes are found the dilaton and the antisymmetric tensor field. Second,
the traditional chiral formulation of gaugino condensation is incorrect in that
it treats the interpolating field $U \simeq \WaWa$ as an ordinary chiral 
superfield of K\"ahler chiral weight $w=2$.  However this is 
inconsistent~\cite{bdqq,bgt,bg96} with the constraint
\be (\DaDa-24R^{\dagger})\WaWa\,-\,(\DbDb-24R)\WbWb
\,=\,\mbox{total derivative,}\ee  
where ${\cal W}^\alpha$ is the Yang-Mills field strength chiral supermultiplet
and the chiral superfield $R$ is an element of the super-Riemann tensor.
On the other hand, the superfield $U$ considered as the chiral projection of the
{\em real} vector superfield $V$ automatically satisfies the constraint
(1.1) with $\WaWa\to U$. Moreover the implementation of the Green-Schwarz
anomaly cancellation mechanism is simpler in the linear multiplet
formulation~\cite{bg96} and much closer in spirit to what happens at the string
level.

As mentioned above, our analysis in \cite{us} only dealt with a pure Yang-Mills
$E_8$  hidden  sector. This was chosen  for the purpose of  illustration of the
method but  has  several  drawbacks  from the point  of view of  phenomenology.
First, the  gauge coupling  blows up  very  close to the  unification scale and
therefore   does  not  allow  for a  large   hierarchy.  Second,  there  are no
moduli-dependent  threshold  corrections and  therefore this  cannot be used to
fix the vacuum expectation values of moduli fields, using for example T-duality
arguments.

A more  realistic  situation  which would  involve   moduli-dependent threshold
corrections,  would be the case of a  hidden sector gauge group being a product
of  simple  groups:  $\G  =  \prod_a  \G_a$. One  immediate   difficulty is the
following:  since  we want to describe  several gaugino  condensates $U_a\simeq
{\WaWa}_a$, we  need to  introduce several  vector superfields  $V_a$. However,
since the theory  has a  single dilaton $\ell$, it  must be identified with the
lowest   component of  $V  =  \sum_a V_a$.  What  should we  do with  the other
components  $\ell_a = V_a|_{\theta = \bar \theta = 0}$?  
We will see  that, in our  description, these are
nonpropagating   degrees  of   freedom  which  actually do  not  appear in the
Lagrangian.  Similarly  only one  antisymmetric  tensor field  (also associated
with $V=\sum_a V_a$) is  dynamical. This allows us  to generalize our  approach
to the case of multicondensates. 

Let us stress
that the goal is very different from the so-called ``racetrack'' 
ideas~\cite{race} where going to the multicondensate case is necessary 
in order to get supersymmetry breaking. Here supersymmetry 
is broken already for a single condensate. Indeed, we will see that the 
picture which emerges from the multicondensate case
(complete with threshold corrections and Green-Schwarz mechanism) 
is very different from the standard ``racetrack'' description:
indeed, the scalar potential is largely dominated by the 
condensate with the largest one-loop beta-function coefficient.

To be more precise,  we generalize in this paper the Lagrangian of~\cite{us} 
to models with 
arbitrary hidden sector gauge groups and with three untwisted (1,1) moduli 
$T^{I}$. We take the K\"ahler potential for the effective theory at the
condensation scale to be:
\bea K &=& k(V) + \sum_Ig^I,\;\;\;\; 
g^I = -\ln(T^I + \bar{T}^I), \;\;\;\; V = \sum_{a=1}^nV_a, \eea
where the $V_a$ are real vector supermultiplets and $n$ is the number of
(asymptotically free) nonabelian gauge groups $\G_a$ in the hidden sector:
\be \G_{\rm hidden} = \prod_{a=1}^n\G_a\otimes U(1)^m. \ee
We will take $\G_{\rm hidden}$ to be a subgroup of $E_8$.

We introduce both gauge condensate superfields $U_a$ and matter
condensate superfields $\Pi^\alpha$ that are nonpropagating:
\be U_a\simeq {\WaWa}_a,\quad \Pi^\alpha\simeq \prod_A\(\Phi^A\)^{n^A_\alpha},
\ee
where ${\cal W}_a$ and $\Phi^A$ are the gauge and matter chiral superfields,
respectively.  The condensate $\Pi^\alpha$ is a chiral superfield of K\"ahler 
chiral weight $w=0$, while the condensate $U_a$ associated with $\G_a$ is a 
chiral superfield of weight $w=2$, and is identified with the chiral projection 
of $V_a$:
\be U_a\,=\,-(\DbDb-8R)V_a, \;\;\;\; \bar{U}_a\,=\,-(\DaDa-8R^{\dagger})V_a.\ee
We are thus introducing $n$ scalar fields $\ell_a = V_a\lowest$. However
only one of these is physical, namely $\ell = \sum_a\ell_a$; the
others do not appear in the effective component Lagrangian constructed below.

The effective Lagrangian for gaugino condensation is constructed and analyzed in
Sections 2--5.  In an appendix we discuss a parallel construction using the 
chiral supermultiplet formulation for the dilaton and unconstrained chiral 
supermultiplets for the gaugino condensates, in order to illustrate the 
differences between the two approaches.  In Section 6 we summarize our results 
and comment on their implications for gauge coupling unification.

\section{Construction of the effective Lagrangian}
\setcounter{equation}{0}
We adopt the following superfield Lagrangian:
\be \Lag_{eff} = \Lag_{KE} + \Lag_{GS} + \Lag_{th} + \Lag_{VY} + \Lag_{pot},\ee
where
\be \Lag_{KE} = \superint\,E \[-2 + f(V)\], \quad k(V) = \ln\,V + g(V),  \ee
is the kinetic energy term for the dilaton, chiral and gravity superfields. The
functions $f(V),g(V)$ parameterize nonperturbative string effects. They are 
related by the condition 
\be
V\frac{\diff g(V)}{\diff V}\,=\,
-V\frac{\diff f(V)}{\diff V}\,+\,f,
\ee which ensures that the Einstein term has canonical form~\cite{us}.
In the classical limit $g=f=0$; we therefore impose the weak coupling boundary
condition:
\be
g(V=0)\,=\,0 \;\;\;\mbox{and}\;\;\; f(V=0)\,=\,0. 
\ee
Two counter terms are introduced to cancel the modular anomaly, namely the
Green-Schwarz term~\cite{gs}:
\be \Lag_{GS} = b\superint\,E V\sum_Ig^I,\quad 
b = {C\over8\pi^2}, \ee
and the term induced by string loop corrections~\cite{dkl}:
\bea \Lag_{th} &=& -\sum_{a,I}{b_a^I\over64\pi^2}\superint\,{E\over R}U_a
\ln\eta^2(T^I) + {\rm h.c.}.\eea
The parameters
\be b^I_a = C - C_a + \sum_A\(1 - 2q^A_I\)C^A_a, \quad C = C_{E_8},\ee
vanish for orbifold compactifications with no $N=2$ supersymmetry 
sector~\cite{ant}. Here $C_a$ and $C_a^A$ are quadratic Casimir operators in 
the adjoint and matter representations, respectively, and $q^A_I$ are the
modular weights of the matter superfields $\Phi^A$ of the underlying hidden
sector theory. The term
\be \Lag_{VY} = 
\sum_a{1\over8}\superint\,{E\over R}U_a\[b'_a\ln(e^{-K/2}U_a/\mu^{3})
 + \sum_\alpha b^\alpha_a\ln\Pi^\alpha\] + {\rm h.c.}, \ee
where $\mu$ is a mass parameter of order one in reduced Planck units (that we
will set to unity hereafter), 
is the generalization to supergravity~\cite{tom,bg89} of the 
Veneziano-Yankielowicz superpotential term, including~\cite{matter} the gauge 
invariant composite matter fields $\Pi^\alpha$ introduced in Eq.(1.4)
(one can also take linear combinations of such gauge invariant monomials that
have the same modular weight).  Finally
\bea \Lag_{pot} &=& {1\over2}\superint\,{E\over R}e^{K/2}W(\Pi^\alpha,T^I) + 
{\rm h.c.} \eea
is a superpotential for the matter condensates that respects the symmetries
of the superpotential $W(\Phi^A,T^I)$ of the underlying field theory.

The coefficients $b$ in (2.8) are dictated by the chiral and conformal
anomalies of the underlying field theory. Under modular transformations, we 
have:
\bea T^{I}&\rightarrow&\frac{aT^{I}-ib}{icT^{I}+d},\;\;\;\;
ad-bc=1,\;\;\;a,b,c,d\;\in\mbox{Z},\nonumber \\
g^I &\to& F^I + \bar{F}^I, \quad F^I = \ln(icT^I + d), \quad
\Phi^A\to e^{-\sum_IF^Iq_I^A}\Phi^A,\nonumber \\
\lambda_a &\to& e^{-{i\over2}\sum_I{\rm Im}F^I}\lambda_a,\quad
\chi^A\to e^{{1\over2}\sum_I(i{\rm Im}F^I - 2q^A_IF^I)}\chi^A, \quad
\theta\to e^{-{i\over2}\sum_I{\rm Im}F^I}\theta,\nonumber \\
U_a &\to& e^{-i\sum_I{\rm Im}F^I}U_a, \quad \Pi^\alpha \to 
e^{-\sum_IF^Iq_I^\alpha}\Pi^\alpha,\quad q^\alpha_I = \sum_A
n^A_\alpha q^A_I .\eea
The field theory loop correction to the effective Yang-Mills Lagrangian from 
orbifold compactification has been determined~\cite{gt,kl} using supersymmetric
regularization procedures that ensure a supersymmetric form for the modular
anomaly.  Matching the variation under (2.10) of that contribution to the
Yang-Mills Lagrangian with the variation of the effective Lagrangian (2.8) we 
require
\be \delta\Lag_{VY} = - {1\over64\pi^2}\sum_{a,I}\int d^4\theta 
{E\over R}U_a\[C_a - \sum_{A,I}C^A_a\(1 - 2q_I^A\)\]F^I + {\rm h.c.}, \ee 
which implies
\be b'_a + \sum_{\alpha,A} b^\alpha_an^A_\alpha q_I^A =
{1\over8\pi^2}\[C_a - \sum_AC^A_a\(1 - 2q^A_I\)\]\;\;\;\; \forall \;\;I. \ee
In the flat space limit where the reduced Planck mass $m_P\to\infty$, under 
a canonical scale transformation 
$$\lambda \to e^{{3\over2}\sigma}\lambda,\;\;\;\; U\to e^{3\sigma}U,\;\;\;\;
\Phi^A\to e^{\sigma}\Phi^A,\;\;\;\;
\Pi^\alpha\to e^{\sum_An^A_\alpha\sigma}\Pi^\alpha,\;\;\;\;
\theta\to e^{-{1\over2}\sigma}\theta,$$ we have the standard
trace anomaly as determined by the $\beta$-functions:
\be\delta\Lag_{eff} = {1\over64\pi^2}\sigma\sum_{a}\int d^4\theta 
{E\over R}U_a\(3C_a - \sum_AC^A_a\) + {\rm h.c.} + O(m_P^{-1}),\ee
which requires
\be 3b'_a + \sum_{\alpha,A}b^\alpha_an^A_\alpha =
{1\over8\pi^2}\(3C_a - \sum_AC^A_a\) + O(m_P^{-1}). \ee 
Eqs. (2.12) and (2.14) 
are solved by~\cite{matter} [up to $O(m_P^{-1})$ corrections]
\be b'_a = {1\over8\pi^2}\(C_a - \sum_AC_a^A\) ,\;\;\;\; 
\sum_{\alpha,A} b^\alpha_an^A_\alpha q_I^A = \sum_A{C^A_a\over4\pi^2}q_I^A,
\quad \sum_{\alpha,A} b^\alpha_an^A_\alpha = \sum_A{C^A_a\over4\pi^2}.  \ee
Note that the above arguments do not completely fix $\Lag_{eff}$ since we can 
{\it a priori} add chiral and modular invariant terms of the form
\be \Delta\Lag = \sum_{a,\alpha}b'_{a\alpha}\superint EV_a\ln
\(e^{\sum_Iq^\alpha_Ig^I}\Pi^\alpha\bar{\Pi}^\alpha\).\ee
For specific choices of the $b'_{a\alpha}$ the matter condensates $\Pi^\alpha$
can be eliminated from the effective Lagrangian.  However the resulting 
component Lagrangian has a linear dependence on the unphysical scalar fields 
$\ell_a - \ell_b$, and their equations of motion impose physically unacceptable 
constraints on the moduli supermultiplets.
To ensure that $\Delta\Lag$ contains the fields $\ell_a$ only through the 
physical combination $\sum_a\ell_a$, we have to impose $b'_{a\alpha} = 
b'_\alpha$ independent of $a$.  If these terms were added the last condition 
in (2.15) would become 
\be \sum_{\alpha,A}b^\alpha_an^A_\alpha + \sum_Ab'_{\alpha}n^A_\alpha
= \sum_A{C^A_a\over4\pi^2}.\ee 
We shall not include such terms here. 

Combining (2.7) with (2.15) gives $
b_a^I = 8\pi^2\(b - b'_a - \sum_\alpha
b^\alpha_aq_I^\alpha\)$. Superspace partial integration gives, for $X$ any
chiral superfield of zero K\"ahler chiral weight:
\be {1\over8}\superint\,{E\over R}U_a\ln X + {\rm h.c.}=
\superint\,E V_a\ln(X\bar{X}) $$ $$
- \pp_m\(\superint\,{E\ln X\over8R}\D_{\dot\alpha}
V_aE^{\dot\alpha m}  + {\rm h.c.}\) , \ee
where $E^{\dot\alpha m}$ is an element of the supervielbein, and the 
total derivative on the right hand side contains the chiral anomaly 
($\propto \pp_mB^m \simeq F^a_{mn}{\tilde F}^{mn}_a $) of the F-term on the 
left hand side.  Then combining the terms (2.2)--(2.9), the ``Yang-Mills''
part of the Lagrangian 
(2.1) can be expressed -- up to a total derivatives that
we drop in the subsequent analysis -- as a modular invariant D-term:
\bea \Lag_{eff} &=& \superint\,E \Bigg( -2 + f(V) + \sum_aV_a\Bigg\{
b'_a\ln(\bar{U}_aU_a/e^gV) + \sum_\alpha
b^\alpha_a\ln\(\Pi_r^\alpha\bar{\Pi}_r^\alpha\)
\nonumber \\ & & \;\; - \sum_I{b_a^I\over8\pi^2}\ln\[\(T^I + \bar{T}^I\)
|\eta^2(T^I)|^2\]\Bigg\}\Bigg) + \Lag_{pot},\eea
where 
\be\Pi_r^\alpha = \prod_A(\Phi_r^A)^{n^A_\alpha} = e^{\sum_Iq^\alpha_Ig^I/2}
\Pi^\alpha,\;\;\;\; \Phi_r^A = e^{\sum_Iq^A_Ig^I/2}\Phi^A, \ee
is a modular invariant field composed of elementary fields that are canonically
normalized in the vacuum.  The interpretation of this result in terms of
renormalization group running will be discussed below.
We have implicitly assumed affine level-one 
compactification.  The generalization to higher affine levels is trivial.  

The construction of the component field Lagrangian obtained from 
(2.19) parallels that given in~\cite{us} for the case $\G=E_8$. Since the
superfield Lagrangian is a sum of F-terms that contain only spinorial
derivatives of the superfield $V_a$, and the Green-Schwarz and kinetic terms 
that contain $V_a$ only through the sum $V$, the unphysical 
scalars $\ell_a$ appear in the component Lagrangian only through the physical 
dilaton $\ell$. The result for the bosonic Lagrangian is
\bea \frac{1}{e}\Lag_{B}\,&=&\,-\,\frac{1}{2}{\cal R}
\,-\,(1+b\ell)\sum_{I}\frac{1}{(t^{I}+\bar{t}^{I})^{2}}
\(\pp^{m}\bar{t}^{I}\,\pp_{\!m}t^{I} - \bar{F}^{I}F^{I}\) \nonumber\\
& &\,-\,\frac{1}{16\dilaton^{2}}\(\dilaton\dg+1\)\[4\(\pp^{m}\!\dilaton\,
\pp_{\!m}\!\dilaton - B^{m}\!B_{m}\) + \bar{u} u - 4e^{K/2}\ell\(W\bar{u} + 
u\bar{W}\)\]\nonumber\\
& &\,+\,\frac{1}{9}\(\dilaton\dg-2 \)\[\bar{M}\!M - b^{m}b_{m} - {3\over4}\lbr
\bar{M}\(\sum_bb'_bu_b - 4We^{K/2}\) + {\rm h.c.}\rbr
\]\nonumber\\
& &\,+\,\frac{1}{8}\sum_a\Bigg\{{f+1\over\ell} + \,b'_a\ln(e^{2-K}\bar{u}_au_a) 
+ \sum_\alpha b^\alpha_a\ln(\pi^\alpha\bar{\pi}^\alpha) \nonumber \\ & & \qquad 
+ \sum_I\[bg^I - {b_a^I\over4\pi^2}\ln|\eta(t^I)|^2\]\Bigg\}
\(F_a - u_a\bar{M} + {\rm h.c.}\)\nonumber\\ 
& &\,-\,\frac{1}{16\dilaton}\sum_a\[b'_a\(\dilaton\dg+1\)\bar{u}u_a -4\ell u_a
\(\sum_\alpha b^\alpha_a{F^\alpha\over\pi^\alpha} + (b'_a-b){F^I\over2{\rm
Re}t^I}\) + {\rm h.c.} \] \nonumber\\
& &\,+\,\frac{i}{2}\sum_a\[b'_a\ln(\frac{u_a}{\bar{u}_a}) + \sum_\alpha 
b^\alpha_a\ln(\frac{\pi^\alpha}{\bar{\pi}^\alpha})\]\nabla^{m}\!B_{m}^a
\,-\,{b\over2}\sum_{I}\frac{\pp^{m}{\rm Im}{t}^{I}}{{\rm Re}{t}^{I}}B_{m}, 
\nonumber \\ & &\,+\,\sum_{I,a}{b^I_a\over16\pi^2}\[\zeta(t^I)\(2iB_a^m\nabla_m
t^I - u_aF^I\) + {\rm h.c.}\]
\nonumber \\ & &\,+\,e^{K/2}\[\sum_IF^I\(W_I + K_IW\) + \sum_\alpha
F^\alpha W_\alpha + {\rm h.c.}\],\eea
where 
\bea  \zeta(t)\,&=&\,{1\over\eta(t)}{\pp\eta(t)\over\pp t}, \quad
\eta(t) = e^{-\pi t/12}\prod_{m=1}^{\infty}\(1-e^{-2m\pi t}\), \nonumber \\
\dilaton\,&=&\,V\lowest,\nonumber\\
\sigma^{m}_{\alpha\dot{\alpha}}B^a_{m}\,&=&\,
\frac{1}{2}[\,\Diff_{\alpha},\Diff_{\dot{\alpha}}\,]V_a\lowest\,+\,
\frac{2}{3}\dilaton_a\sigma^{m}_{\alpha\dot{\alpha}} b_{m},
\quad B^m = \sum_aB^m_a,\nonumber\\
u_a\,&=&\,U_a\lowest\,=\,-(\bar{\Diff}^{2}-8R)V_a\lowest,
\quad u = \sum_au_a,\nonumber\\
\bar{u}_a\,&=&\,\bar{U}_a\lowest\,=\,-(\Diff^{2}-8R^{\dagger})V_a\lowest,
\quad \bar{u} = \sum_a\bar{u}_a,\nonumber \\
-4F^a\,&=&\,\Diff^{2}U_a\lowest, \;\;\;\;
-4\bar{F}^a\,=\,\bar{\Diff}^{2}\bar{U}_a\lowest, 
\quad F_U = \sum_aF^a,\nonumber\\
\pi^\alpha\,&=&\,\Pi^\alpha\lowest\,
\quad \bar{\pi}^\alpha\,=\,\bar{\Pi}^\alpha\lowest\,\nonumber \\
-4F^\alpha\,&=&\,\Diff^{2}\Pi^\alpha\lowest, \;\;\;\;
-4\bar{F}^\alpha\,=\,\bar{\Diff}^{2}\bar{\Pi}^\alpha\lowest, \nonumber \\
t^{I}\,&=&\,T^{I}\lowest,\;\;\;\;
-4F^{I}\,=\,\Diff^{2}T^{I}\lowest,\nonumber\\
\bar{t}^{I}\,&=&\,\bar{T}^{I}\lowest,\;\;\;\;
-4\bar{F}^{I}\,=\,\bar{\Diff}^{2}\bar{T}^{I}\lowest,\eea 
$b_m$ and $M = (\bar{M})^{\dag} = -6R\lowest$ 
are auxiliary components
of the supergravity multiplet~\cite{bggm}. For $n=1,\;u_a=u,\;etc.,$
(2.21) reduces to the result of~\cite{us}.  

The equations of motion for the auxiliary fields $b_m,
M,F^I,F^a+\bar{F}^a$ and $F^\alpha$ give, respectively,:
\bea b_m\,&=&\,0, \;\;\;\; 
M\,=\,\frac{\, 3\,}{\, 4\,}\,\(\sum_ab'_au_a - 
4We^{K/2}\) ,\nonumber\\
F^{I}\,&=&\,{{\rm Re}t^I\over2(1 + b\ell)}\lbr\sum_a\bar{u}_a\[(b - b'_a) +
{b_a^I\over2\pi^2}\zeta(\bar{t}^I){\rm Re}t^I\]- 4
e^{K/2}\(2{\rm Re}t^I\bar{W}_I - \bar{W}\)\rbr, \nonumber \\ 
\bar{u}_au_a\,&=&\,\frac{\ell}{e^{2}}e^{g\,-\,({f+1})/{b'_a\dilaton}-\sum_I
b^I_ag^I/8\pi^2b'_a}\prod_I|\eta(t^I)|^{b^I_a/2\pi^2b'_a}
\prod_\alpha(\pi^\alpha_r\bar{\pi}^\alpha_r)^{-b^\alpha_a/b'_a},
\quad \pi^\alpha_r = \Pi^\alpha_r\lowest, \nonumber \\ 
0\,&=&\,\sum_ab^\alpha_au_a\,+\,4\pi^\alpha e^{K/2}W_\alpha \quad\forall \;\;
\alpha.\eea
Using these, the Lagrangian (2.17) reduces to
\bea \frac{1}{e}\Lag_{B}\,&=&\,-\,\frac{1}{2}{\cal R}
\,-\,(1+b\ell)\sum_I\frac{\pp^{m}\bar{t}^{I}\,\pp_{\!m}t^{I}}
{(t^{I}+\bar{t}^{I})^{2}}\,-\,\frac{1}{4\dilaton^{2}}\(\dilaton\dg+1\)
\(\pp^{m}\!\dilaton\,\pp_{\!m}\!\dilaton - B^{m}\!B_{m}\)  \nonumber\\
& &\,-\,\sum_a\(b'_a\omega_a + \sum_\alpha b^\alpha_a\phi^\alpha\)
\nabla^{m}\!B_{m}^a\,-\,{b\over2}\sum_{I}
\frac{\pp^{m}{\rm Im}{t}^{I}}{{\rm Re}{t}^{I}}B_{m} \nonumber \\ & &
+\,i\sum_{I,a}{b^I_a\over8\pi^2}\[\zeta(t^I)B_a^m\nabla_mt^I - {\rm h.c.}\] - V,
\nonumber\\ V\,&=&\,\frac{\(\dilaton\dg+1\)}{16\dilaton^{2}}\lbr\bar{u}u +\ell
\[\bar{u}\(\sum_ab'_au_a - 4e^{K/2}W\)  + {\rm h.c.} \]\rbr\nonumber \\ 
& &\,+\,{1\over 16(1+b\ell)}\sum_I\left|\sum_au_a\(b-b'_a + {b^I_a\over2\pi^2}
\zeta(t^I){\rm Re}t^I\)  - 4e^{K/2}\(2{\rm Re}t^IW_I - W\)\right|^2
\nonumber\\ & &\,+\,\frac{1}{16}\(\dilaton\dg-2 \)\left|\sum_bb'_bu_b
- 4We^{K/2}\right|^2 ,\eea
where we have introduced the notation
\bea u_a &=& \rho_ae^{i\omega_a}, \quad \pi^\alpha = \eta^\alpha
e^{i\phi^\alpha}, \eea
and \bea 2\phi^\alpha &=& 
-i\ln\(\sum_a b^\alpha_au_a\bar{W}_\alpha\over\sum_a b^\alpha_a\bar{u}_a
W_\alpha\) \quad {\rm if} \;\; W_\alpha\ne 0.
\eea   
To go further we have to be more specific.  Assume\footnote{For, {\it e.g.,} 
$\G= E_6\otimes SU(3),$ we take 
$\Pi\simeq (27)^3$ of $E_6$ or $(3)^3$ of $SU(3)$.} that for fixed $\alpha,\;
b^\alpha_a\ne 0$ for only one value of $a$.  
For example, we allow no representations $(n,m)$ with both $n$ and $m\ne 1$ 
under $\G_a\otimes\G_b$. Then $u_a= 0$ unless $W_\alpha\ne 0
$ for every $\alpha$ with $b^\alpha_a\ne 0$. We therefore assume 
that $b^\alpha_a \ne 0$ only if $W_\alpha\ne 0$.  

Since the $\Pi^\alpha$ are gauge invariant operators, 
we may take $W$ linear in $\Pi$:
\be W(\Pi,T) = \sum_\alpha W_\alpha(T)\Pi^\alpha, \;\;\;\;  
W_\alpha(T) =  c_\alpha\prod_I[\eta(T^I)]^{2(q^\alpha_I - 1)},\ee
where $\eta(T)$ is the Dedekind function. If there are gauge singlets $M^i$
with modular weights $q^i_I$, then the constants
$c_\alpha$ are replaced by modular invariant functions:
$$c_\alpha\to w_\alpha(M,T) = c_\alpha\prod_i(M^i)^{n^\alpha_i}
\prod_I[\eta(T^I)]^{2n^\alpha_iq^i_I}.$$  In addition if some $M^i$ have gauge 
invariant couplings to vector-like representations of the gauge group 
$$W(\Phi,T,M)\ni c_{iAB}M^i\Phi^A\Phi^B
\prod_I[\eta(T^I)]^{2(q_I^A + q_I^B + q_I^i)},$$ one has to
introduce condensates $\Pi^{AB}\simeq \Phi^A\Phi^B$ of dimension two, and
corresponding terms in the effective superpotential: $$W(\Pi,T,M)\ni 
c_{iAB}M^i\Pi^{AB}\prod_I[\eta(T^I)]^{2(q_I^A + q_I^B + q_I^i)}.$$  Since the
$M^i$ are unconfined, they cannot be absorbed into the composite fields $\Pi$.
The case with only vector-like representations has been considered
in~\cite{matter}.  To simplify the present discussion, we ignore this
type of coupling and assume that the composite operators that are invariant 
under the gauge symmetry (as well as possible discrete global symmetries) are 
at least trilinear in the nonsinglets under the confined gauge group.  We
further assume that there are no continuous global symmetries--such as a flavor
$SU(n)_R\otimes SU(n)_L$ whose anomaly structure has to be 
considered~\cite{matter}.  With these assumptions the equations of motion 
(2.23) give, using $\sum_\alpha b^\alpha_a q^\alpha_I + {b_a^I/8\pi^2} 
= b-b'_a$,
\bea \rho^2_a &=& e^{-2b_a'/b_a}e^Ke^{-(1+f)/b_a\ell - b\sum_Ig^I/b_a}
\prod_I|\eta(t^I)|^{4(b-b_a)/b_a}
\prod_\alpha|b^\alpha_a/4c_\alpha|^{-2b_a^\alpha/b_a}, \nonumber \\
\pi_r^\alpha &=& - e^{-{1\over2}[k+ \sum_I(1-q^\alpha_I)g^I]}
{b^\alpha_a\over4W_\alpha}u_a, 
\quad  b_a \equiv b'_a + \sum_\alpha b^\alpha_a. \eea
Note that promoting the second equation above to a superfield relation, and 
substituting the expression on the right hand side for $\Pi$ in (2.19) gives
\bea \Lag_{eff} &=& \superint\,E \Bigg( -2 + f(V) + \sum_aV_a\Bigg\{
b_a\ln(\bar{U}_aU_a/e^gV) \nonumber \\ & & \quad - \sum_\alpha b_a^\alpha
\ln\(e^{\sum_Ig^I(1 - q^\alpha_I)}\left|4W_\alpha/b_a^\alpha\right|^2\)
\nonumber \\ & & \quad - \sum_I{b_a^I\over8\pi^2}\ln\[\(T^I + \bar{T}^I\)
|\eta^2(T^I)|^2\]\Bigg\}\Bigg) + \Lag_{pot}. \eea
It is instructive to compare this result with the effective Yang-Mills 
Lagrangian 
found~\cite{gt,kl} by matching field theory and string loop calculations.
Making the identifications $V\to L, \;U_a\to{\WaWa}_a$, the effective Lagrangian
at scale $\mu$ obtained from those results can be written:
\bea \Lag^{YM}_{eff}(\mu) &=& \superint\,E \Bigg( -2 + f(V) + \sum_aV_a\Bigg\{
{1\over 8\pi^2}\(C_a - {1\over3}\sum_AC^A_a\)
\ln\[{\mu^6_sg_{s}^{-4}\over\mu^6g_a(\mu)^{-4}}\] \nonumber \\ & & \quad
- {1\over4\pi^2}\sum_AC^A_a\ln\[g_s^{2\over3}Z_A(\mu_s)/g_a^{2\over3}(\mu)
Z_A(\mu)\] \nonumber \\ & & \quad 
- \sum_I{b_a^I\over8\pi^2}\ln\[\(T^I + \bar{T}^I\)
|\eta^2(T^I)|^2\]\Bigg\}\Bigg), \eea with $\mu^2_s \sim g^2_s \sim \ell$
in the string perturbative limit, $f(V) = g(V) = 0$.  The first term in
brackets in 
(2.29) can be identified with the corresponding term (2.30) provided
\be \sum_\alpha b^\alpha_a = {1\over12\pi^2}\sum_AC^A_a,\;\;\;\; b_a =
{1\over8\pi^2}\(C_a - {1\over3}\sum_AC^A_a\). \ee
In fact, this constraint follows from (2.15) if the $\Pi^\alpha$ are all of 
dimension three, which is consistent with the fact that only dimension-three 
operators survive in the superpotential in the limit $m_P\to \infty$. Then
$b_a$ is proportional to the $\beta$-function
for $\G_a$, and $\rho_a \simeq <\bar{\lambda}_a\lambda_a>$ has the
expected exponential suppression factor for small coupling. 
In the absence of nonperturbative corrections to the K\"ahler potential 
[$f(V) = g(V) = 0$], $\langle V\lowest\rangle = <\ell> = g^2_s = \mu_s^2$ is 
the string scale in reduced Planck units and also the gauge coupling at that
scale~\cite{gt,kl}.  Therefore the argument of the log:
\be \left<\({\bar{U}_aU_a\over V}\)^{1\over3}\right> = 
{\left<\(\bar{\lambda}_a\lambda_a\)^{1\over3}\right>\over g_s^{2\over3}} = 
{\left<\(\bar{\lambda}_a\lambda_a\)^{1\over3}\right>
\over\mu_s^2g_s^{-{4\over3}}} \ee
gives the exact two-loop result for the coefficient of 
$C_a$ in the renormalization group running from the string scale to the
appropriate condensate scale~\cite{bg91,gt,kl}.  
The relation between $<\pi^\alpha>$ and $<u_a>$, and hence the appearance of 
the gaugino condensate as the effective infra-red cut-off for massless matter 
loops, is related to the Konishi anomaly~\cite{kon}.  The matter loop 
contributions have additional two-loop corrections involving matter wave 
function renormalization~\cite{barb,Ant,gjs}:
\bea {\pp\ln Z_A(\mu)\over\pp\ln\mu^2} &=& -
{1\over 32\pi^2}\Bigg[\ell e^g\sum_{BC}e^{\sum_Ig^I\(1- q^A_I - q^B_I - 
q^C_I\)}Z^{-1}_A(\mu)Z^{-1}_B(\mu)Z^{-1}_C(\mu)|W_{ABC}|^2\nonumber \\ & &
- 4\sum_ag^2_a(\mu)C_2^a(R_A) \Bigg] + O(g^4) + O(\Phi^2),\eea
where $C_2^a(R_A) = ({\rm dim}\G_a/{\rm dim}R_A)C_a^A, \;R_A$ 
is the representation of $\G_a$ on $\Phi_A$.  The boundary condition on 
$Z_A$ is~\cite{gt} $Z_A(\mu_s) = (1 - p_A\ell)^{-1}$ where $p_A$ is the 
coefficient of $e^{\sum_Iq^A_Ig^I}|\Phi^A|^2$ in the Green-Schwarz counter 
term in the underlying field theory: $V = \sum_Ig^I + 
p_Ae^{\sum_Iq^A_Ig^I}|\Phi^A|^2 + O(|\Phi^A|^4)$.  The second line of (2.29) 
can be interpreted as a rough parameterization of the second line of (2.30).

In the following analysis, we retain only dimension three operators in the
superpotential, and do not include any unconfined matter superfields in the
effective condensate Lagrangian.  
The potential takes the form
\bea V &=& {1\over16\ell^2}\sum_{a,b}\rho_a\rho_b\cos\omega_{ab}R_{ab}(t^I),
\quad \omega_{ab} = \omega_a - \omega_b, \nonumber 
\\ R_{ab} &=& \(\ell\dg+1\)\(1 + b_a\ell\)\(1 + b_b\ell\) - 3\ell^2b_ab_b
+ {\ell^2\over(1 + b\ell)}\sum_Id_a(t^I)d_b(\bar{t}^I), 
\nonumber \\ d_a(t^I) &=& b - b'_a + {b^I_a\over2\pi^2}\zeta(t^I){\rm Re}t^I
- \sum_\alpha b^\alpha_a\[1 - 4(q^\alpha_I - 1){\rm Re}t^I\zeta(t^I)\] 
\nonumber \\ &=& \(b - b_a\)\(1 + 4\zeta(t^I){\rm Re}t^I\).\eea
Note that $d_a(t^I)\propto F^I$ vanishes at the self-dual point $t^I=1,\;
\zeta(t^I) = -1/4,\;\eta(t^I)\approx .77$.  For 
Re$t^I \stackrel{\textstyle{>}}{\sim}1$ we have, to a very good
approximation, $\zeta(t^I)\approx -\pi/12,\;\eta(t^I)\approx e^{-\pi t/12}$.
Note that also $\rho_a$ -- and hence the potential $V$ -- vanishes in the 
limits of large and small radii; from (2.28) we have
\bea \lim_{t^I\to \infty}\rho_a^2 &\sim  & (2{\rm Re}t^I)^{(b-b_a)/b_a}
e^{-\pi(b-b_a){\rm Re}t^I/3b_a}, \nonumber \\
 \lim_{t^I\to 0}\rho_a^2 & \sim & (2{\rm Re}t^I)^{(b_a -b)/b_a} 
e^{-\pi (b-b_a)/3b_a{\rm Re}t^I}, \eea
where the second line follows from the first by the duality invariance of
$\rho^2_a$. 
So there is potentially a ``runaway moduli problem''.   However, as shown in
Section 4, the moduli are stabilized at a physically acceptable vacuum, namely
the self-dual point.

\section{The axion content of the effective theory}
\setcounter{equation}{0}
Next we consider the axion states of the effective theory.  If all 
$W_\alpha\ne 0$, the equations of motion for $\omega_a$ obtained from 
(2.24) read:
\be {\pp\Lag\over\pp\omega_a} = -b'_a\nabla^mB^a_m - {1\over2}
\sum_{\alpha,b}b^\alpha_b\({b^\alpha_a u_a\over\sum_c b^\alpha_cu_c} + 
{\rm h.c.}\)\nabla^mB^b_m - {\pp V\over\pp\omega_a}=0. \ee
These give, in particular,
\bea \sum_a{\pp\Lag\over\pp\omega_a} &=& - \sum_ab_a\nabla^mB^a_m = 0. \eea
The one-forms $B_m^a$ are {\it a priori} dual to 3-forms:
\be B_m^a = {1 \over 2} \epsilon_{mnpq} \left( {1 \over 3!4} 
\Gamma^{npq}_a + \partial^n b_a^{pq}\right),\ee
where $\Gamma^{npq}_a$ and $b_a^{pq}$ are 3-form and 2-form potentials,
respectively; (3.3) assures the constraints 
(1.1) for $\WaWa\to U_a$; explicitly
\be
(\DaDa-24R^{\dagger})U_a\,-\,(\DbDb-24R)\bar{U}_a \,=\,-2i{}^* \Phi_a = 
-{2i \over 3!} \epsilon_{mnpq} \partial^m \Gamma^{npq}_a = -16i\nabla^mB_m^a.
\ee
We obtain 
\be -b'_a{}^*\Phi_a - {1\over2}\sum_{\alpha,b}b^\alpha_b
\({b^\alpha_a u_a\over\sum_c b^\alpha_cu_c} + {\rm h.c.}\){}^*\Phi_b =
8{\pp V\over\pp\omega_a}, \quad \sum_ab_a{}^*\Phi_a = 0. \ee
If $\Gamma^{npq}\ne 0,\;b^{pq}$ can be removed by a gauge
transformation $\Gamma^{npq}\to \Gamma^{npq} + \pp^{[n}\Lambda^{pq]
}$.  Thus
\be B_m^a = {1\over 2nb_a} \epsilon_{mnpq}\partial^{n}\tb^{pq} +
{1 \over 3!8} \epsilon_{mnpq}\Gamma^{npq}_a,\quad \sum_ab_a\Gamma^{npq}_a = 0, 
\quad \tb^{pq} = \sum_ab_ab^{pq}_a,\ee
and we have the additional equations of motion:
\be {\delta\over\delta\tb_{pq}}\Lag_B = 0, \;\;\;\;
\({1\over b_a}{\delta\over\delta\Gamma^{a}_{npq}} - {1\over b_b}
{\delta\over\delta\Gamma^{b}_{npq}}\)\Lag_B = 0,\quad
{\delta\over\delta\phi}\Lag_B \equiv {\pp\Lag_B\over\pp\phi} - 
\nabla^m\({\pp\Lag_B\over\pp(\nabla^m\phi)}\), \ee
which are equivalent, respectively, to
\be \epsilon_{mnpq}\sum_a{1\over b_a}\nabla^n{\delta\over\delta B_a^m}\Lag_B = 
0, \quad \({1\over b_a}{\delta\over\delta B_a^m} - 
{1\over b_b}{\delta\over\delta B_b^m}\)\Lag_B = 0, \ee
with  
\bea {1\over e}{\delta\over\delta B^a_m}\Lag_B &=& 
{\(\ell\dg+1\)\over2\ell^2}B^m + b'_a\pp^m\omega_a  + {1\over2}
\sum_{\alpha,b}b^\alpha_a\({b^\alpha_b u_b\over\sum_c b^\alpha_cu_c}
+ {\rm h.c.}\)\pp^m\omega_b \nonumber \\ & &
+ \sum_\alpha b^\alpha_a\[\pp^m\ell{\pp\phi^\alpha\over\pp\ell} +
\sum_I\(\pp^m t^I{\pp\phi^\alpha\over\pp t^I} + {\rm h.c.}\)\]\nonumber \\ & &
+ i\sum_{a,I}{b^I_a\over8\pi^2}\[\zeta(t^I)\pp^mt^I - {\rm h.c.}\]
- {b\over2}\sum_I{\pp^m{\rm Im}t^I\over {\rm Re}t^I} .\eea
Combining these with (3.1) and the equations of motion for $\ell$ and $t^I$, one
can eliminate $B^a_m$ to obtain the equations of motion for an equivalent
scalar-axion Lagrangian, with a massless axion dual to $\tb_{mn}$.

Again, these equations simplify considerably if we assume that for fixed 
$\alpha,\;b^\alpha_a\ne 0$ for only one value of $a$.  In this case (3.1) 
reduces to 
\be \nabla^mB^a_m = - {1\over b_a}{\pp V\over\pp\omega_a}, \ee
and we have 
\be {\pp\phi^\alpha\over\pp\ell} = 0,\quad {\pp\phi^\alpha\over\pp t^I} =
i\zeta(t^I)\(q^\alpha_I -1\),\ee
if we restrict the potential to terms of 
dimension three with no gauge singlets $M^i$. Using $\sum_\alpha
b^\alpha_a\(q_I^\alpha - 1\) + b^I_a/8\pi^2 = b-b_a$ gives:
\bea{1\over e}{\delta\over\delta B^a_m}\Lag_B &=& {\(\ell\dg+1\)\over2\ell^2}
B^m + b_a\pp^m\omega_a + i\sum_I\lbr\pp^mt^I\[\zeta(t^I)\(b-b_a\) 
+ {b\over4{\rm Re}t^I}\] - {\rm h.c.}\rbr
\nonumber \eea \be \approx {\(\ell\dg+1\)\over2\ell^2}B^m + 
b_a\pp^m\omega_a + \sum_I\pp^m{\rm Im}t^I\[\(b - b_a\){\pi\over6}
- {b\over2{\rm Re}t^I}\] ,\ee
where the last line corresponds to the approximation 
$\zeta(t^I)\approx-\pi/12$. In the following we illustrate these equations using
specific cases.

\subsection{Single condensate} 
As in~\cite{us} there is an axion $\omega = \omega_a + 
(\pi/6)(b/b_a - 1)\sum_I{\rm Im}t^I$ that has no potential, and, setting
\be B_a^m = {1\over2}\epsilon^{mnpq}\pp_nb_{pq} = -
{2\ell^2\over\(\ell\dg + 1\)}\(b_a\pp^m\omega - {b\over2}\sum_I{\pp^m
{\rm Im}t^I\over{\rm Re}t^I}\),\ee
the equations of motion derived from (2.24) are equivalent to 
those of the effective scalar Lagrangian:
\be \frac{1}{e}\Lag_{B}\,=\,-\,\frac{1}{2}{\cal R}
\,-\,(1+b\ell)\sum_I\frac{\pp^{m}\bar{t}^{I}\,\pp_{\!m}t^{I}}
{(t^{I}+\bar{t}^{I})^{2}}\,-\,\frac{1}{4\dilaton^{2}}\(\dilaton\dg+1\)
\pp^{m}\!\dilaton\,\pp_{\!m}\!\dilaton  - V(\ell,t^I,\bar{t}^I) $$ $$ 
\,-\,{\ell^2\over\(\ell\dg + 1\)}\(b_a\pp^m\omega - {b\over2}\sum_I{\pp^m
{\rm Im}t^I\over{\rm Re}t^I}\)\(b_a\pp_m\omega - {b\over2}\sum_I{\pp_m
{\rm Im}t^I\over{\rm Re}t^I}\) . \ee

\subsection{Two condensates: $b_1\ne b_2$} 
Making the approximation $\eta(t)\approx e^{-\pi t/12}$, the Lagrangian 
(2.24) can be written
\bea \frac{1}{e}\Lag_{B}\,&=&\,-\,\frac{1}{2}{\cal R}
\,-\,(1+b\ell)\sum_I\frac{\pp^{m}\bar{t}^{I}\,\pp_{\!m}t^{I}}
{(t^{I}+\bar{t}^{I})^{2}}\,-\,\frac{1}{4\dilaton^{2}}\(\dilaton\dg+1\)
\(\pp^{m}\!\dilaton\,\pp_{\!m}\!\dilaton - B^{m}\!B_{m}\) \nonumber \\ & &
\,-\,\omega\nabla^{m}\!\tB_{m} -
\omega'\nabla^{m}\!B_{m}\,-\,{b\over2}\sum_{I}\frac{\pp^{m}
{\rm Im}{t}^{I}}{{\rm Re}{t}^{I}}B_{m} - V, \eea
where
\bea  \omega &=& {b_1\omega_1 - b_2\omega_2\over b_1 - b_2}  - 
{\pi\over 6}\sum_I{\rm Im}t^I, \quad \omega' = -{\omega_{12} \over\beta}
+ {b\pi\over 6}\sum_{I}{\rm Im}{t}^{I},\nonumber \\ 
\beta &=& {b_1 - b_2\over b_1b_2}, \quad \tB^m = \sum_a b_aB^m_a . \eea
We have 
\bea \omega_1 &=& \omega + {\pi\over6}\sum_I{\rm Im}t^I + {1\over b_1}
\(\omega'- {b\pi\over6}\sum_I{\rm Im}t^I\), \nonumber \\
\omega_2 &=& \omega + {\pi\over6}\sum_I{\rm Im}t^I + {1\over b_2}\(\omega'
- {b\pi\over6}\sum_I{\rm Im}t^I\), \nonumber \\
{\pp V\over\pp\omega_1} &=& -{\pp V\over\pp\omega_2} = 
{\pp V\over\pp\omega_{12}}.\eea
Then taking $\omega,\omega'$ and $t^I$ as independent variables, the equations
of motion for $\omega,\omega'$ are
\bea \nabla^m\tB_m &=& 0,\quad \tB_m = 
{1\over 2}\epsilon_{mnpq}\partial^{n}\tb^{pq}, \nonumber \\
\nabla^mB_m &=& {1\over8}{}^*\Phi = \beta{\pp V\over\pp\omega_{12}}, 
\quad B_m = {1 \over 3!8}\epsilon_{mnpq}\Gamma^{npq}. \eea 
Substituting the first of these into the Lagrangian (3.15), we see that the
axion $\omega$ and the three-form $\tB_m$ drop out because they appear only
linearly in the Lagrangian; hence they play the role of Lagrange multipliers.
The equation of motion for $\tb_{mn}$ implies the constraint on the phase 
$\omega$:
\be \nabla_m\pp^m\omega = 0.\ee
The equations of motion for Im$t^I$ and $\Gamma_{mnp}$ read:
\bea 0 &=& \nabla_m\[\(1 + b\ell\){\pp^m{\rm Im}{t}^{I}\over2
\({\rm Re}{t}^{I}\)^2} + {b\over2{\rm Re}{t}^{I}}B^{m}\] 
-  i\({\pp V\over\pp t^I} - {\rm h.c.}\) - 
{b\pi\over48}{}^*\Phi, \nonumber \\ 
0 &=& {\(\ell\dg+1\)\over2\ell^2}B^m + \pp^m\omega' - {b\over2}
\sum_I{\pp^m{\rm Im}t^I\over{\rm Re}t^I},\eea
and the equivalent scalar Lagrangian is
\bea \frac{1}{e}\Lag_{B}\,&=&\,-\,\frac{1}{2}{\cal R}
\,-\,(1+b\ell)\sum_I\frac{\pp^{m}\bar{t}^{I}\,\pp_{\!m}t^{I}}
{(t^{I}+\bar{t}^{I})^{2}}\,-\,\frac{1}{4\dilaton^{2}}\(\dilaton\dg+1\)
\pp^{m}\!\dilaton\,\pp_{\!m}\!\dilaton  \nonumber \\ & & 
\,-\,{\ell^2\over\(\ell\dg + 1\)}\(\pp^m\omega' -
{b\over2}\sum_I{\pp^m{\rm Im}t^I\over{\rm Re}t^I}\)\(\pp_m
\omega' - {b\over2}\sum_I{\pp_m{\rm Im}t^I\over{\rm Re}t^I}\) \nonumber \\ & &
\,-\,V(\ell,t^I,\bar{t}^I,\omega_{12}) . \eea
As in Subsection A, there is a single dynamical axion $\omega'$ -- or, 
via a duality
transformation, ${}^*\Phi$ -- but there is now a potential for the axion.

\subsection{General case} 
We introduce $n$ linearly independent vectors 
$\tB_m,B_m,\hB^i_m,\;i= 1\ldots n-2$, and decompose the $B^m_a$ as
\bea B^m_a &=& a_a\tB^m + c_aB^m + \sum_id^i_a\hB^m_i, \quad \hB^m_i =
\sum_a e_i^a B^m_a. \eea
Then
\be \sum_a\[b_a\omega_a + (b-b_a){\pi\over6}\sum_I{\rm Im}t^I\]
\nabla_mB^m_a = \omega\nabla_m\tB^m + \omega'\nabla_mB^m
+ \sum_i\omega^i\nabla_m\hB^m_i, $$ $$
\omega_a = \omega + {\pi\over6}\sum_I{\rm Im}t^I + {1\over b_a}\(\omega' - 
{b\pi\over6}\sum_I{\rm Im}t^I\) + \sum_i{e_i^a\over b_a}\omega^i,\ee
and the Lagrangian can be written as in (3.15) with an additional term:
\be {1\over e}\Lag_B \to {1\over e}\Lag_B - \sum_i\omega^i\nabla_m\hB^m_i, \ee
The equations of motion for the phases $\omega$ are:
\bea \nabla_m\tB^m &=& -
{\pp V\over\pp\omega} = -\sum_a{\pp V\over\pp\omega_{a}} = 0,\nonumber \\
\nabla_mB^m &=& - {\pp V\over\pp\omega'} = -
\sum_a{1\over b_a}{\pp V\over\pp\omega_{a}} = {1\over2}\sum_{ab}\beta_{ab}
{\pp V\over\pp\omega_{ab}} ={1\over8}{}^*\Phi, 
\quad \beta_{ab}\equiv {b_a - b_b\over b_ab_b}
\nonumber \\ \nabla_m\hB^m_i &=& -{\pp V\over\pp\omega^i} = -
\sum_a{e_i^ a\over b_a}{\pp V\over\pp\omega_{a}} = {1\over8}{}^*\Phi_i,\eea
and the equations for $\Gamma_{mnp}^i= 8\epsilon_{mnpq}\hB^q_i$ give
$\pp^m\omega^i = 0$. Hence
\be \omega_{ab} = - \beta_{ab}\(\omega' - 
{b\pi\over6}\sum_I{\rm Im}t^I\) + \theta_{ab}, \;\;\;\; \theta_{ab} =
{\rm constant}. \ee
Thus as in the two-condensate case of Subsection B, 
there is one dynamical axion with a
potential. The dual scalar Lagrangian is the same as (3.21), with 
$V = V(\ell,t^I,\bar{t}^I,\omega_{ab})$.

\section{The effective potential}
\setcounter{equation}{0}
The potential (2.34) can be written in the form 
\bea V &=& {1\over16\ell^2}\(v_1 - v_2 + v_3\),\nonumber \\ v_1 &=& 
\(1+\ell\dg\)\left|\sum_a\(1+b_a\ell\)u_a\right|^2,  \quad 
v_2 = 3\ell^2\left|\sum_ab_au_a\right|^2,\nonumber \\ 
v_3 &=& {\ell^2\over(1+b\ell)}\sum_I\left|\sum_ad_a(t^I)u_a\right|^2
= 4\ell^2\(1+b\ell\)\sum_I\left|{F^I\over{\rm Re}t^I}\right|^2. \eea
In the strong coupling limit
\be \lim_{\ell\to\infty}V = \(\ell\dg - 2\)\left|\sum_ab_au_a\right|^2 ,\ee
giving the same condition on the functions $f,g$ as in~\cite{us} to assure
boundedness of the potential.  Note however that if $v_1 = v_3 = 0$ has a 
solution with $v_2\ne 0$, the vacuum energy is always negative.  $v_3 = 0$ is 
solved by $t^I=1$, {\it i.e.} the self-dual point. As explained below
this is the only nontrivial minimum if the cosmological constant is fine-tuned
to vanish. 
In the case of two condensates, there is no solution to  $v_1 = 0,\;v_2\ne 0$, 
for $f\ge0$, and the cosmological constant can be fine-tuned to vanish,
as will be illustrated below in a toy example.  More generally, 
the potential is dominated by the condensate with the largest one-loop
$\beta$-function coefficient, so the general case is qualitatively very similar 
to the single condensate case, and it appears that positivity of the potential 
can always be imposed. Otherwise, one would have to
appeal to another source of supersymmetry breaking to cancel the
cosmological constant, such as a fundamental 3-form potential~\cite{three} whose
field strength is dual to the constant that has been previously introduced in
the superpotential~\cite{drsw}, and/or an anomalous $U(1)$ gauge 
symmetry~\cite{u1}.

In the following we study $Z_3$-inspired toy models with $E_6$ and/or $SU(3)$ 
gauge groups in the hidden sector, and $3N_f$ matter superfields~\cite{iban}
in the fundamental representation $f$. Asymptotic freedom requires $N_{27}\le3$ 
and $N_3 \le 5$.  For a true $Z_3$ orbifold there are no moduli-dependent
threshold corrections: $b^I_a=0$.  In this case universal anomaly cancellation 
determines the average value of the matter modular weights in these toy models 
as: $<2q_I^{27}-1> = 2/N_{27},\;<2q_I^{3}-1> = 18/N_3$. 
In some models Wilson line breaking of the hidden sector $E_8$ generates 
vector-like representations that could 
acquire masses above the condensation scale, so that the universal anomaly 
cancellation sum rule is not saturated by light states alone.  In this case 
the $q^\alpha_I$ no longer drop out of the equations, so some of the above 
formulae would be slightly modified.  In addition, 
one would have to include threshold effects~\cite{kl,kamran}, unless the 
masses of the heavy states are pushed to the string scale.   Here we assume 
for simplicity that the sum rule is saturated by the light states. 
Denoting the fundamental matter fields by $\Phi_f^{I\alpha},\;\alpha = 1,
\ldots,N_f$, the matter condensates can be constructed as 
$$\Pi_f^\alpha = \prod_{I=1}^3\Phi_f^{I\alpha}, \quad b^\alpha_{E_6} = 
{3\over4\pi^2}, \quad b^\alpha_{SU(3)} = {1\over8\pi^2}, $$
where gauge indices have been suppressed.

In the analysis of the models described below, we assume--for obvious
phenomenological reasons--that the vacuum energy vanishes at the minimum
$<V> = 0$. 
Thus we solve the equations 
\be V = {\pp V\over\pp x} = 0, \quad x = \ell,t^I,\omega_a.\ee
For $x = \ell,t^I$, we have
\bea {\pp \rho_a\over\pp x} &=& {1\over2}\(A_x + {1\over b_a}B_x\)\rho_a, 
\quad B_\ell = {(1+\ell\dg)\over\ell^2},\quad B_I = {b\over2{\rm Re}t^I}\[1 + 4
\zeta(t^I){\rm Re}t^I\], \nonumber \\
{\pp V\over\pp x} &=& \(A_x -{2\over\ell}\delta_{x\ell}\)V + 
{1\over16\ell^2}\sum_{ab}\rho_a\rho_b\cos\omega_{ab}\({B_x\over b_a}R_{ab} + 
{\pp\over \pp x}R_{ab}\) 
\nonumber \\ &=& {1\over16\ell^2}\sum_{ab}\rho_a\rho_b\cos\omega_{ab}
\({B_x\over n}\sum_{c}\beta_{ca}R_{ab} + {\pp\over\pp x}R_{ab}\) 
\nonumber \\ & & \qquad + 
\(A_x -{2\over\ell}\delta_{x\ell} + {B_x\over n}\sum_{a}{1\over b_a}\)V,\eea
where $\beta_{ab}$ is defined in (3.25).
By assumption, the last term in (4.4) vanishes in the vacuum. Note that the
self-dual point, $d_a(t^I) = B_I = 0,\;t^I = 1,$ is always a solution to the 
minimization equations for $t^I$. It is the only solution for the single
condensate case.  For the multicondensate case, if we restrict 
our analysis to the (relatively) weak coupling region, $\ell < 1/b_-$, 
where $b_-$ is the smallest $\beta$-function constant, the potential is 
dominated by the condensate with the largest $\beta$-function coefficient
$b_+:\;V\approx
\rho^2_+R_{++}/16\ell^2$.  Moreover, since $\pi b/3b_a > 1$, the potential is
always dominated by this term for Re$t^{I}>1$ [{\it c.f.} Eq. (2.35)], so 
the only 
minimum for Re$t^I>1$ is Re$t^I \to\infty,\;\rho_a\to 0$.  By duality the only 
minimum for Re$t^I <1$ is Re$t^I \to 0,\;\rho_a\to 0$, so the self-dual point 
is the only nontrivial solution.  Since our potential is always dominated by one
condensate, the picture is very different from the ``race-track'' models studied
previously~\cite{race}.

At a self-dual point with $V=0$, we have 
\bea {\pp^2 V\over\pp(t^I)^2} &\approx& {1\over32\ell^2}\sum_{ab}\rho_a\rho_b
\cos\omega_{ab}\({\pi^2\over9}{\ell^2\over(1+b\ell)}(b-b_a)(b-b_b) 
- {b\pi\over6n}\sum_{c}\beta_{ca}R_{ab}\)\nonumber \\ 
&\approx& {\rho^2_+\over32}\({\pi^2\over9}{(b-b_+)^2\over(1+b\ell)}
 - {b\pi\over6n\ell^2}\sum_{c}\beta_{c+}R_{++}\) . \eea
Positivity of the potential requires $R_{++}\ge 0$, and $\beta_{c+}\le 0$ by
definition, so the extremum at a self-dual point with $V=0,\;\rho_+\ne0$
is a true minimum.  In practice, the last term is negligible, and the 
normalized moduli squared mass is \be m^2_{t^I} \approx
\left< {\pi^2 \rho^2_+\over36}{(b-b_+)^2\over(1+b\ell)^2}\right>.\ee

\subsection{Single condensate with matter}
In this case $\beta_{ab} = 0$, and the minimization equations for $t^I$
require 
$${\pp\over\pp t^I}\left|1 + 4\zeta(t^I){\rm Re}t^I\right|^2 = 0,$$ which is 
solved by $1 + 4\zeta(t^I){\rm Re}t^I = 0,\; t^I=1.$  
Then $v_3=F^I=0,$ and the potential is qualitatively the same as
in the $E_8$ case~\cite{us}--except for the fact that the moduli are fixed.
(Note however that if $\beta_{ab}= 0$ one can choose the $b'_{a\alpha}$ in 
(2.16) such that the matter composites drop out of the effective Lagrangian; 
then $R_{aa}$ is independent of the moduli which remain undetermined.) The
quantitative difference from the $E_8$ case is the value of the
$\beta$-function coefficient: $b_{E_6} = {\(12 - 3N_{27}\)/8\pi^2},\;b_{SU(3)}= 
{\(6-N_3\)/16\pi^2}.$  As in~\cite{us} we take the nonperturbative 
contribution to the dilaton K\"ahler potential to be of the form~\cite{bd} 
$f = Ae^{-B/\ell}$ or~\cite{shenk} $f = Ae^{-B/\sqrt{\ell}},$ and fine tune the 
constant $A$ to get a vanishing cosmological constant.

Attention has been drawn to the leading
correction for small coupling that is of the form $f = Ae^{-B/\sqrt{V}}$.  If 
we restrict $f$ to this form we have to require a rather large value for the
coefficient: $A\simeq 40$ to cancel the cosmological constant.  On the other 
hand the important feature of $f$ here is its behaviour in the strong coupling 
limit; if $f$ contains terms of the form $Ae^{-B/V^{n\over 2}}$ the strong
coupling limit will be dominated by the term with the largest value of $n$.
In the numerical analysis we take $f = Ae^{-B/V}$; adding to this a term of 
the form $f = A'e^{-B'/\sqrt{V}}$ will not significantly affect the analysis.
We find that the $vev$ of $\ell$ is insensitive to the content of the hidden
sector; it is completely determined by string nonperturbative effects,
provided a potential for $\ell$ is generated by the strongly coupled hidden
Yang-Mills sector.  More specifically, taking $f = Ae^{-B/V}$ we find that 
$<V> = 0$ requires $A\approx e^2 \approx 7.4$, and the dilaton is stabilized 
at a value $<\ell> \approx B/2$.  Taking $B = 1$  gives $<\ell> \approx0.5,
\; <f(\ell)> \approx 1,$ and the squared gauge coupling at the string scale is
$g_s^2 = <2\ell/(1+f)> \approx 0.5.$  If instead we use $f = Ae^{-B/\sqrt{V}}$, 
the corresponding numbers are $A\approx2e^3 \approx 40,\; <\ell> \approx 
B^2/9,\; g_s^2 \approx 2B^2/27.$  From now on we take $f = Ae^{-1/V}$. 

The potential for $\G_a = E_6,\; N_{27} = 1$, is plotted in Figures 1--3.  
Fig. 1 shows the
potential in the $\ell,\ln t$ plane, where we have set $t^I=t$, Im$t=0$; with 
this choice of variables the $t$-duality invariance of the potential is
manifest.  Fig. 2 shows the potential for $\ell$ at the self-dual point
$t^I=1$, and Fig. 3 shows the potential for $\ln t$ with $\ell$ fixed at its
$vev$.  The qualitative features of the potential are independent of the content
of the hidden sector.  Fixing $A$ in each case by the condition $<V>=0$, we
find for $\G_a = E_6$
\be A = \cases{7.324\cr7.359\cr7.381\cr}, \quad <\ell> = \cases{0.502\cr0.501
\cr0.500\cr}\approx g_s^2,\quad{\rm for}\quad N_{27} = \cases{1\cr2\cr3\cr}.\ee
For $\G_a = SU(3),\;N_3 = 1$, we find $A= 7.383,\;<\ell> = 0.500\approx
g^2_s$.  
As discussed in Section 5, the scale of supersymmetry breaking in this case is 
far too small, and decreases with increasing $N_3$.
 
\subsection{Two condensates} 
We have 
\bea {\pp V\over\pp \omega_1} &=& - {\pp V\over\pp \omega_2} = 
- \rho_1\rho_2R_{12}\sin\omega_{12},\nonumber \eea \be
\sum_{abc}\beta_{ca}\rho_a\rho_bR_{ab}\cos\omega_{ab} =
\beta_{21}\(\rho_1^2R_{11} - \rho_2^2R_{22}\).\ee
Minimization with respect to $\omega_{1}$ requires either $<\sin\omega_{12}>= 
0$ or $<R_{12}>= 0$.  Identifying $b_1=b_+,\;b_2=b_-$, positivity of the
potential requires $R_{11}\ge 0$, which in turn implies $R_{12}>0,$ so the
extrema in $\omega$ are at $\sin\omega_{12}=0$, with $\cos\omega_{12} = - 1(+1)$
corresponding to minima (maxima):
\be {\pp^2 V\over\pp\omega_{12}^2} = - \rho_1\rho_2R_{12}\cos\omega_{12},\quad
m^2_{\omega_{12}} = \left< {3b^2_+ \beta^2\over 2(1 + b_+ \ell)^2} 
\rho_1\rho_2R_{12} \right> . \ee
There is also a small Im$t^I$-$\omega_{12}$ mixing. Note that
while in contrast to the single condensate case, the dynamical axion is no
longer massless, its mass is exponentially suppressed relative to the 
gravitino mass by a factor $\sim \sqrt{\rho_2/\rho_1}$.  We do not expect this
feature to persist when kinetic terms are introduced for the condensate fields.

For $\G = E_6\otimes SU(3)$, the potential is dominated by the $E_6$
condensate, and the results are the same as in (4.7).  The only other gauge
groups in the restricted set considered here that are subgroups of $E_8$ are
$\G = [SU(3)]^{n},\; n\le 4$; these cannot generate sufficient supersymmetry
breaking.

\section{Supersymmetry breaking}
\setcounter{equation}{0}
The pattern and scale of supersymmetry breaking are determined by the
$vev$'s of the $F$-components of the chiral superfields.  From the equations 
of motion for $\pi^\alpha$ and $\rho_a$ we obtain, at the self-dual point
$<F^I>=0$:
\bea \langle F^\alpha \rangle &=& {(\ell\dg +1)\over4\ell^2 b_a}\pi^\alpha
\(\bar{u} + \ell\sum_b b_b \bar{u}_b\) \approx{3b_+^2\over4b_a}\pi^\alpha
\bar{u}_+\(1 + \ell b_+\)^{-1}, \quad b^\alpha_a\ne0,\nonumber \\
\langle F^a + \bar{F}^a \rangle &=& {1\over 4 \ell^2 b_a}
(\ell\dg +1)(1+\ell b_a)
\[ u_a \( \bar{u} + \ell \sum_b b_b \bar{u}_b \) + {\rm h.c.}\] \nonumber \\ &&
\approx {3 b^2_+ \over 4b_a}{1+\ell b_a \over 1+\ell b_+}
\( u_a \bar u_+ + \bar u_a u_+ \),  \eea
where the approximations on the right hand sides are exact for a single
condensate.   The dominant contribution is from the condensate with the largest
$\beta$-function coefficient:
\be \langle F^+ + \bar{F}^+ \rangle = 
 {3\rho^2_+b_+\over2}.\ee
The fact that the $F^I$ vanish in the vacuum is a desirable feature for 
phenomenology.  Nonuniversal squark and slepton masses that could induce
unacceptably large flavor-changing neutral currents are thereby avoided.
However this feature might be modified in
the presence of moduli-dependent threshold effects $\sim\ln(\mu^2)$ where
$\mu^2 = <e^{\sum_Iq^i_Ig^I}|M^i|^2>$ is a modular invariant squared mass and
$M^i$ is a gauge singlet with modular weights $q^I_i$.

Another important parameter for soft supersymmetry breaking in the
observable sector is the gravitino mass. The derivation of the 
gravitino part of the Lagrangian again parallels the construction in~\cite{us}.
The gravitino mass $m_{\tilde G}$ is determined by the term
\bea \Lag_{mass}(\psi) &=& -\frac{1}{8}\psi^m\sigma_{mn}\psi^n\sum_a\bar{u}_a
\Bigg\{{f+1\over\ell} + \,b'_a\ln(e^{2-K}\bar{u}_au_a) 
+ \sum_\alpha b^\alpha_a\ln(\pi^\alpha\bar{\pi}^\alpha) \nonumber \\ & & 
+ \sum_I\[bg^I - {b_a^I\over4\pi^2}\ln|\eta(t^I)|^2\]\Bigg\}
- e^{K/2}\bar{W}\psi^m\sigma_{mn}\psi^n + {\rm h.c.},\eea
giving, when the equations of motion (2.23) are imposed,
\be m_{\tilde G} = {1\over3}\langle|M|\rangle = 
\frac{\, 1}{\, 4\,}\,\langle|\sum_ab'_au_a - 4e^{K/2}W|\rangle= 
\frac{\, 1}{\, 4\,}\,\langle|\sum_ab_au_a|\rangle \approx
{1\over 4}b_+\langle\rho_+\rangle. \ee

The scale of supersymmetry breaking is governed by the $vev$ (2.28) of 
the condensate with the largest $\beta$-function coefficient.  This includes the
usual suppression factor $<\rho_a> \propto e^{-1/b_ag_s^2}$, where $g_s^2 =
<2\ell/(1+f)>$ is the effective squared coupling constant at the string
scale.  
However there
are other important parameters that determine the scale of the hierarchy
between the supersymmetry breaking scale and the Planck scale.  The dependence
on the moduli provides a second exponential suppression factor:
\be <\rho_a> \propto <\prod_I |\eta(t^I)|^{2(b - b_a)/b_a}> = 
|\eta(1)|^{6(b - b_a)/b_a} \approx e^{-\pi(b - b_a)/2b_a} .\ee
On the other hand, the numerical factor
$\prod_\alpha|b^\alpha_a/4c_\alpha|^{-b^\alpha_a/b_a}$ gives
an exponential enhancement if $c_\alpha\sim 1$.  This is the largest numerical
uncertainty in our analysis.  {\it A priori}, $c_\alpha$ is related to the
Yukawa couplings for matter in the hidden sector.  However, there is an
arbitrary normalization factor in the definition of $\Pi^\alpha$.  If the
hidden sector Yukawa couplings were known, it might be possible to estimate 
$c_\alpha$ by a matching condition for the $vev$'s of the second lines of 
(2.29) and (2.30). In our numerical analysis we have set $c_\alpha = 1$.
Then if the hidden gauge group with the largest condensate is $\G_+= E_6$ with
$3N_{27}$ matter chiral superfields in the fundamental representation, 
we obtain 
\be m_{\tilde G} = \cases{1.1\times 10^{-9}\cr3.3\times 10^{-11}
\cr1.65\times 10^{-15}\cr} \quad {\rm for} \quad N_{27} = \cases{1\cr2\cr3\cr},
\ee
in reduced Planck units.  For $\G_+= SU(3)$ with three matter chiral fields in
the fundamental representation, we obtain an unacceptably large gauge hierarchy:
$m_{\tilde G} = 2.2\times 10^{-32};\;m_{\tilde G}$ decreases rapidly as $N_{3}$
increases, {\it i.e.} as the $\beta$-function coefficient decreases.

\section{Concluding remarks}
\setcounter{equation}{0}
In the class of models studied here, the introduction of a parameterization for
nonperturbative contributions to the K\"ahler potential for the dilaton 
generically allows a stable
vacuum at a nontrivial, phenomenologically acceptable point in the 
dilaton/moduli space.  In particular, when we impose the constraint that the 
cosmological constant vanishes, we find that in the linear multiplet 
formulation, the moduli $t^I$ are stabilized at the
self-dual point, and their associated auxiliary fields vanish in the vacuum,
which implies the phenomenologically desirable feature of universal soft
supersymmetry breaking parameters. As shown in the Appendix, these
features do not survive in the parallel construction starting from the chiral
multiplet formalism because of the explicit $s$-dependence of the
superpotential.  They may also be modified in the linear multiplet formalism in
the presence of moduli-dependent intermediate-scale threshold effects.  
However the case with
no such threshold corrections serves to illustrate the difference between the
two approaches.  We have argued that the linear multiplet approach more 
faithfully respects the physics of the underlying strongly coupled Yang-Mills
theory.  

A salient feature of our formalism is that there is little 
qualitative difference between a single
condensate and a multi-condensate scenario.  For several condensates with equal
(or very similar) $\beta$-functions, the potential reduces to that of the single
condensate case, except that there may be flat
directions. If $b_1=b_2=\cdots b_k$, then at the self-dual point
$\rho_a/\rho_1=\zeta_a =$ constant and the potential vanishes identically in the
direction $\sum_{a=1}^k\zeta_ae^{i\omega_a}=0,\;\rho_{a>k} = 0.$  This always
has a solution if $\zeta_a = 1$, in which case the flat direction preserves
supersymmetry and there is no barrier between this solution and the interesting,
supersymmetry breaking solution.
For different $\beta$-functions, the potential is dominated by
the condensate(s) with the largest $\beta$-function coefficient, and the
result is essentially the same as in the single condensate case, except that a 
small mass is generated for the dynamical
axion.  In all cases nonperturbative corrections to the dilaton K\"ahler
potential are required to stabilize the dilaton.  This picture is very 
different from previously studied ``racetrack'' models~\cite{race} where 
dilaton stabilization is achieved through cancellations among different 
condensates with similar $\beta$-functions.
The qualitative difference
between an $E_8$ hidden sector and one with a product gauge group is the
presence of matter; in the $E_8$ case the potential is independent of
the moduli, which therefore remain undetermined in the classical vacuum of the
effective condensate theory. 

As discussed previously~\cite{bdqq,bgt,us}, kinetic energy terms for the
condensate fields $\rho_a,\omega_a,$ as well as an axion mass comparable to the
condensation scale, can be generated by including a dependence 
of the K\"ahler potential $k$ (and correspondingly the function $f$) on the
variables $U_a,\bar{U}_a$.  Terms of the form $V^{-2n}\sum_a\(U_a\bar{U}_a\)^n$
and $V^{-2n}\(U\bar{U}\)^n$ are generated both by classical string 
corrections~\cite{Ant2} and by field theory loop corrections~\cite{gjs}.
Note that once the condensate fields are integrated out these induce, by virtue
of their $vev$'s (2.14), ``nonperturbative'' corrections to the K\"ahler
potential for the dilaton, of the type discussed by Banks and Dine~\cite{bd}.
However in the single condensate case~\cite{yy} it was found that these terms 
are insufficient to stabilize the dilaton, and one must appeal instead to 
string nonperturbative effects.\footnote{It can be shown that the static model 
of~\cite{us} is indeed the low energy limit of the dynamical model 
of~\cite{yy}.}  We expect the same conclusion to hold in the 
multicondensate case.  If this is so, the interpretation of contributions to the
K\"ahler potential of the form $f = Ae^{-B/V}$ as arising from field theoretic 
corrections to our our static model may be questionable.  We therefore adopt
the point of view that the unknown function $f$ parameterizes string
nonperturbative corrections.  

In the static models studied here, cancellation of the cosmological 
constant by string nonperturbative corrections alone requires
that they are significant at the vacuum: $<f(\ell)>\approx <2\ell>\approx 1$. 
This has 
implications\footnote{Other gauge-dependent threshold corrections~\cite{kk} 
have recently been found.} for phenomenological analyses~\cite{unif} of
gauge coupling unification.  Including nonperturbative corrections
to the K\"ahler potential for the linear multiplet $L$, {\it i.e.}, taking
$k(L) = \ln L + g(L)$ with $f(L)$ related to $g(L)$ as in (2.3) with $V\to L$, 
the two-loop boundary condition~\cite{gt} on the $\overline{MS}$ gauge 
couplings now reads (for affine level one):
\bea g_a^{-2}(\mu_s) &=& g^{-2}_s + {C_a\over8\pi^2}\lbr g(\ell) + \ln\[f(\ell)
+1\] - \ln 2\rbr \nonumber \\ & & - {1\over16\pi^2}\sum_Ib_a^I
\ln\[\(t^I + \bar{t}^I\)|\eta^2(t^I)|^2\],\nonumber \\
g^{-2}_s &=& {f+1\over2\ell}, \quad \mu^2_s = \ell e^{g-1} = 
{1\over2}e^{g-1}(f+1)g^2_s .\eea
Note that the tree coupling of the effective field theory is now 
$2g_s^{-2} = \left<{(f+1)/\ell}\right>,$ and integration over the condensate
fields with $vev$'s given by 
(2.28) gives corrections to the K\"ahler potential
for $\ell$ of the form~\cite{bd} $\sim e^{-n/b_ag^2_s} = e^{-(f+1)/b_a\ell}$, 
when kinetic terms for $U_a,\bar{U}_a$ are included.  On the other hand, we
expect string nonperturbative effects~\cite{shenk} to be $\sim
e^{-n\pi/\sqrt{\ell}}$
since the linear supermultiplet containing the 3-form $d^{[n}b^{pq]}$ is the
fundamental field in string compactifications (as opposed {\it e.g.} to 5-brane
compactification~\cite{brane}, in which the dilaton is in a chiral multiplet and
the moduli are in linear multiplets).
If one performs a duality transformation in the usual way~\cite{bggml}
{\it via} a Lagrange multiplier $S + \bar{S}$: 
$$\Lag = \superint E\[-2 + f(L) + {1\over3}\(L + \Omega\)\(S + \bar{S}\)\],$$
where $\Omega$ is the Chern-Simons superfield, $L$ is unconstrained and $S$ is 
chiral, the equations of motion for $L$ give precisely $S + \bar{S} = 
(f+1)/L$, so that Re$s$ is always the tree-level inverse squared 
coupling constant in the chiral formulation of the effective field theory.
Including the Green-Schwarz term and loop corrections in the chiral 
formulation~\cite{kl} again gives (6.1).

\vskip .8cm
\noindent {\bf Acknowledgements}
\vskip .5cm
PB and MKG would like to acknowledge support from the Miller Institute for Basic
Research in Science, and also the Aspen Center for Physics where part of this 
work was completed.  This work was supported in part by the Director, Office of 
Energy Research, Office of High Energy and Nuclear Physics, Division of 
High Energy Physics of the U.S. Department of Energy under Contract 
DE-AC03-76SF00098 and in part by the National Science Foundation under 
grant PHY-95-14797.

\vskip .3in
\appendix
\def\ksubsection{\Alph{subsection}}
\def\theequation{\ksubsection.\arabic{equation}} 
\catcode`\@=11
\subsection{Appendix: Chiral multiplet formulation}
\setcounter{equation}{0} 

There has been interest in the question as to whether the linear and chiral
multiplet formulations are equivalent at the quantum level.  They are presumably
equivalent in the sense that we may perform a duality transformation at the
superfield level on the Lagrangian (2.1) so as to recast it entirely in terms of
chiral supermultiplets;  the resulting effective Lagrangian is apt to be rather
complicated.  The more practical question that we address in this appendix is 
the extent to which the above results can be reproduced if one takes as a 
starting point the usual chiral supermultiplet formalism for the dilaton with 
the gaugino condensates represented by unconstrained chiral supermultiplets, 
and na\"{\i}vely generalizes the methods commonly used in this context.

In the chiral multiplet formulation, the Green-Schwarz term appears as a 
correction to the K\"ahler potential, which we take to be 
\be K(S,T^I) = \ln(L) + {\tilde g}(L) + \sum_Ig^I, \quad L^{-1} = S+\bar{S} - b
\sum_Ig^I, \ee
where ${\tilde g}$ is the correction from nonperturbative string effects.  
Modular invariance of the Yang-Mills Lagrangian at the quantum level is 
assured by the transformation property of $S$ under (2.10):
\be S \to S + b\sum_IF^I,\ee
and modular covariance of the K\"ahler potential [$K\to K + \sum_I(F^I +
\bar{F}^I)$] requires that it depend on $S$ only through the real superfield
$L$. We introduce static condensate superfields $\Pi^\alpha,U_a$ as before,
but now the superfield \be U_a = e^{K/2}H_a^3 \ee does not satisfy the 
constraint (3.4) because $H_a$ is taken to be an unconstrained chiral 
superfield.\footnote{This 
is probably where the departure from the approach of Section 2 is
the most sensitive.  The correct procedure -- which is not the one usually
followed -- would be to use a 3-form supermultiplet description~\cite{three}.}  
We construct the superpotential in analogy to (2.1), using the standard 
Veneziano-Yankielowicz approach:
\be W_{tot} =  W_{cond} + W(\Pi), \ee
where $W(\Pi)$ is the same as in (2.27), and
\bea W_{cond} &=& W_C + W_{VY} + W_{th}, \quad W_C = {1\over4}S\sum_aH_a^3,
\nonumber \\ W_{VY} &=& {1\over4}\sum_aH_a^3\(3b'_a\ln H_a + 
\sum_\alpha b^\alpha_a\ln\Pi^{\alpha}\), \nonumber \\ 
W_{th} &=& {1\over4}\sum_{a,I}{b^I_a\over8\pi^2}H_a^3\ln[\eta^2(T^I)],\eea
where $W_C$ represents the classical contribution.  $H_a^3$ transforms in the
same way as $U_a$ under rigid chiral and conformal transformations,  and the
anomaly matching conditions give the same constraints on the $b$'s as in
Section 2.  Then it is straightforward to check that under the modular 
transformation (2.10) with $H_a\to e^{-\sum_IF^I/3},$ we have $W_{cond}\to 
e^{-\sum_IF^I/3}W_{cond},$ as required for modular invariance of the Lagrangian.
Summing the various contributions, the superpotential for $H_a$ can be written
in the form 
\be W_{cond} = {1\over4}\sum_a b'_aH_a^3\ln\lbr e^{S/b'_a}H^3_a\prod_\alpha
(\Pi^\alpha)^{b^\alpha_a/b'_a}\prod_I[\eta(T^I)]^{-b^I_a/4\pi^2b'_a}\rbr. \ee
The bosonic Lagrangian takes the standard form:
\bea \Lag_B &=& -{1\over2}{\cal R} - {1\over3}M\bar{M} + K_{i\m}\(F^i
\bar{F}^{\m} - \pp_\mu z^i\pp^\mu z^{\m}\) \nonumber \\ & & 
+ e^{K/2}\[F^i\(W_i + K_iW\) - \bar{M}W + {\rm h.c.}\],\eea
where $Z^i = S,T^I,H_a,\Pi^\alpha,\;z^i=Z^I\lowest$.  In our static model 
$K_{i\m},K_i = 0$ for $Z^i,Z^m = H_a,\Pi^\alpha$, and the equations of motion 
for $F^i$ give $W_i = 0$ for these fields.  This determines the chiral 
superfields $H_a,\Pi^\alpha$ as holomorphic functions of $S,T^I$.  Making the 
same restrictions on $W(\Pi)$ and the $b^\alpha_a$ as in Section 2, we obtain:
\bea H^3_a &=& e^{(2n+1)i\pi(b'_a - b_a)/b_a - b'_a/b_a}e^{-S/b_a}\prod_I
[\eta(T^I)]^{2(b - b_a)/b_a}\prod_\alpha\left|{b_a^\alpha/4c_\alpha}
\right|^{-b_a^\alpha/b_a},  \nonumber \\
\Pi^\alpha &=& -{b_a^\alpha\over4c_\alpha}H_a^3\prod_I[\eta(T^I)]^{-2(q^\alpha_I
- 1)},\quad b_a^\alpha \ne 0. \eea
As in (2.28), the correct dependence of the gaugino condensates on the gauge
coupling constant $<({\rm Re}s)^{-{1\over2}}>,\; s = S\lowest$, is recovered.  
Note however that in contrast to (2.28) the gaugino condensate phases are 
quantized once Im$s$ is fixed at its $vev$.  Using these results gives
\be W_{tot} = W(S,T^I) = - {1\over4} \sum_a  b_aH_a^3.\ee
The effective potential is determined in the standard way after eliminating the
remaining auxiliary fields through their equations of motion:
\bea M &=& - 3e^{K/2}W,\quad \bar{F}^{\m} = - e^{K/2}K^{i\m}\(W_i + K_iW\), 
\quad Z^i = S,T^I, \nonumber \\
V(s,t^I,\bar{t}^I) &=& e^{K}\[K^{i\m}\(W_i + K_iW\)\(\bar{W}_{\m} + 
K_{\m}\bar{W}\) - 3|W|^2\].\eea
The inverse K\"ahler metric for the K\"ahler potential (A.1) is 
\bea K^{I\bar{J}} &=& {4({\rm Re}t^I)^2\over(1-bK_s)}\delta^{IJ}, \quad
K^{I\bar{s}} = - {2b{\rm Re}t^I\over(1-bK_s)}, \nonumber \\
K^{s\bar{s}} &=& {1 - bK_s + 3b^2K_{s\bar{s}}\over K_{s\bar{s}}(1-bK_s)},\eea
and the potential reduces to 
\bea V &=& {e^K\over1-bK_s}\bigg\{K_{s\bar{s}}^{-1}\(1-bK_s + 3b^2K_{s\bar{s}}\)
|W_s + K_sW|^2 + 4\sum_I\({\rm Re}t^I\)^2|W_I + K_IW|^2 \nonumber \\ & &
- 2b\[\(\bar{W}_s + K_s\bar{W}\)\sum_I{\rm Re}t^I\(W_I + K_IW\) + {\rm h.c.}\]
\bigg\} - 3e^K|W|^2.\eea
We have
\bea - 2{\rm Re}t^I\(W_I + K_IW\) &=& -\sum_a{1\over4b_a}\[1-bK_s 
- {b-b_a\over b_a}{\rm Re}t^I\zeta(t^I)\]H^3_a, \nonumber \\
W_s + K_sW &=& \sum_a{1\over4b^2_a}\(1 - K_sb_a\)H^3_a ,\eea
and the potential can be written in the form
\be V = {e^K\over16(1-bK_s)}\sum_{ab}|h_ah_b|^3\cos\omega_{ab}R_{ab}, \ee
where now $\omega_a$ is the phase of $h^3_a= H^3_a\lowest,\;\omega_{ab}$ 
is defined as before, and 
\bea R_{ab} &=& b_ab_bf_{ab}(\ell) + (b-b_a)(b-b_b)\sum_I|1 + 
4{\rm Re}t^I\zeta(t^I)|^2, \quad \ell = L\lowest,\nonumber \\ 
f_{ab}(\ell) &=& (1 - bK_s)\[{(1- b_aK_s)(1- b_aK_s)\over b_ab_bK_{s\bar{s}}} 
- 3\]. \eea
In the absence of nonperturbative effects $K_s = -\ell,\;K_{s\bar{s}} = \ell^2,
\;f_{ab} \to - 2b\ell$ as $\ell\to \infty$, and the potential is unstable in 
the strong
coupling direction, as expected.  A positive definite potential requires that
$f_{++}(\ell)$ be postive semi-definite where, as before, $b_+$ is the largest 
$b_a$.
Note that the perturbative expression for $f_{aa}(\ell)$ is negative for 
$b_a\ell>1.4$, while in the linear multiplet formalism, the corresponding 
expression is negative only for $b_a\ell> 2.4$, so nonperturbative effects are
required to be more important in the chiral multiplet formulation.  If there 
is only one condensate, the self-dual point for the moduli is again a
minimum, but $<F^I> \ne 0$.  In the general case, the minimization equations 
for the moduli read
\bea {\pp V\over\pp t^I} &=& {e^K\over16(1-bK_s)}
\sum_{ab}|h_ah_b|^3\cos\omega_{ab}
\({2b\over n}\zeta(t^I)\sum_{c}\beta_{ca}R_{ab} + {\pp\over\pp t^I}R_{ab}\) 
\nonumber \\ & & + \(A + {2b\over n}\zeta(t^I)\sum_{a}{1\over b_a}\)V , \eea
where $\beta_{ab}$ is defined as in (3.25).
Again imposing $<V>= 0$, the minimum is shifted slightly away from the self dual
point if some $\beta_{ab}\ne 0$.

The effective Lagrangian in the linear multiplet formalism -- like the
string and field theory loop-corrected Yang-Mills Lagrangian~\cite{gt,kl} --
depends only on the variables $t^I$ and the modular invariant field $\ell$, so
the Lagrangian is invariant under modular transformations on the $t^I$ alone.
In contrast, the effective Lagrangian in the standard chiral multiplet approach
has an explicit $s$-dependence which accounts for the fact that the self-dual
point is not the minimum.  The standard chiral construction forces a holomorphic
coefficient for the interpolating superfield for the Yang Mills composite
superfield $U\simeq \WaWa$, and hence cannot faithfully reflect
nonholomorphic contributions from the Green-Schwarz term and field theory loop
corrections.  The last point can be evaded by incorporating these 
renormalization effects 
in the K\"ahler potential~\cite{eta,bg96,kamran} rather than in the 
superpotential, in which case it is also possible to recover invariance under
continuous infinitesimal S-duality rotations in the weak coupling limit.  Again
this a property of the Yang-Mills Lagrangian and the linear multiplet
formulation of condensation, but not of the chiral multiplet 
formulation.\footnote{This can again be traced~\cite{bg96} 
to the fact that the condensate 
superfield (A.3) does not satisfy the constraint (3.4).}
However, in this last approach the relation of the effective Lagrangian for
condensation to the underlying Yang-Mills Lagrangian is much less transparent.
We emphasize that we do not claim that there is no effective chiral Lagrangian
dual to that of Section 2, with the same physics.  However a straightforward
approach based on the chiral multiplet formalism leads to different physics,
in particular the nonvanishing of the moduli F-terms in the vacuum, which has
implications for flavor-changing neutral currents.

\vskip 1in
\hspace{1.6in} FIGURE CAPTIONS

\vskip 0.5in
\mbox{Fig. 1:} The scalar potential $V$ (in reduced Planck units)
is plotted versus $\ell$ and $\ln t$.

\vskip 0.5in
\mbox{Fig. 2:} The scalar potential $V$ (in reduced Planck units)
is plotted versus $\ell$ with $t^{I}=1$ (the self-dual point).

\vskip 0.5in
\mbox{Fig. 3:} The scalar potential $V$ (in reduced Planck units)
is plotted versus $\ln t$ with $\,\ell\,=\,\langle\,\ell\,\rangle$.

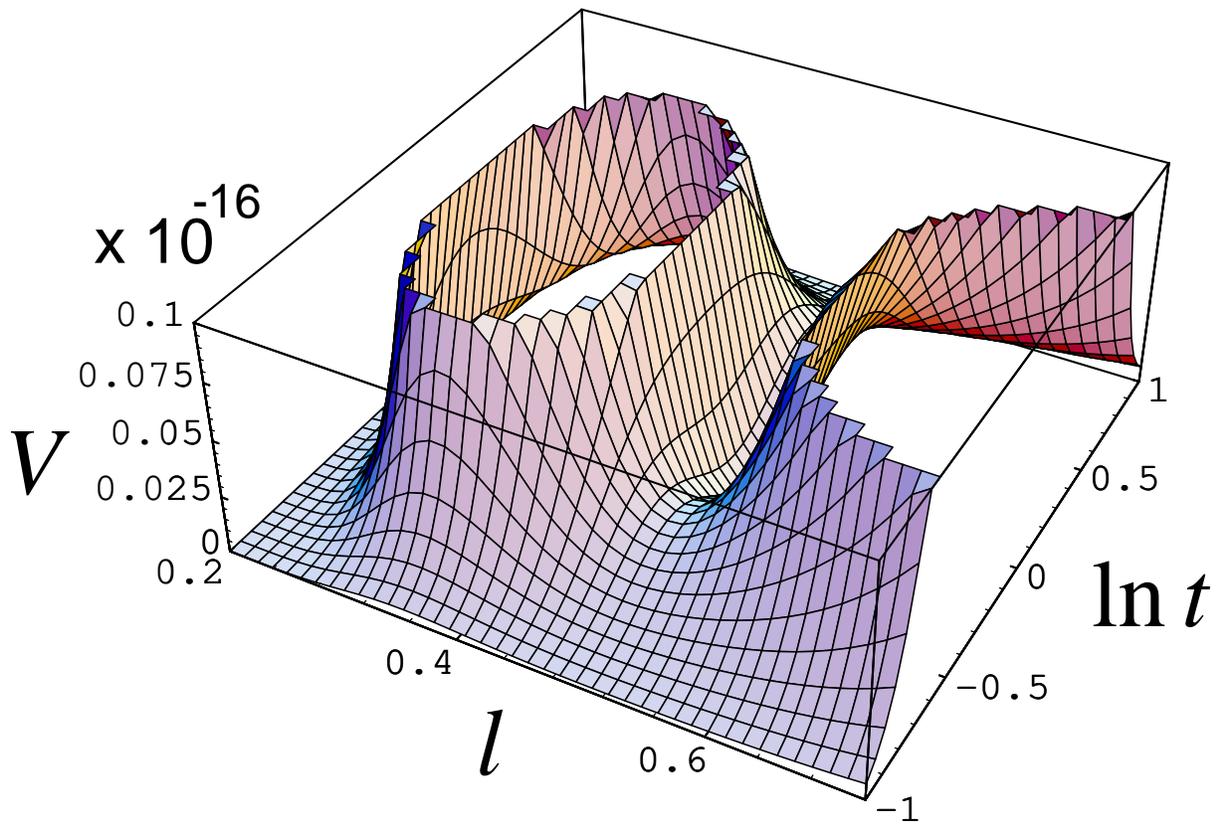
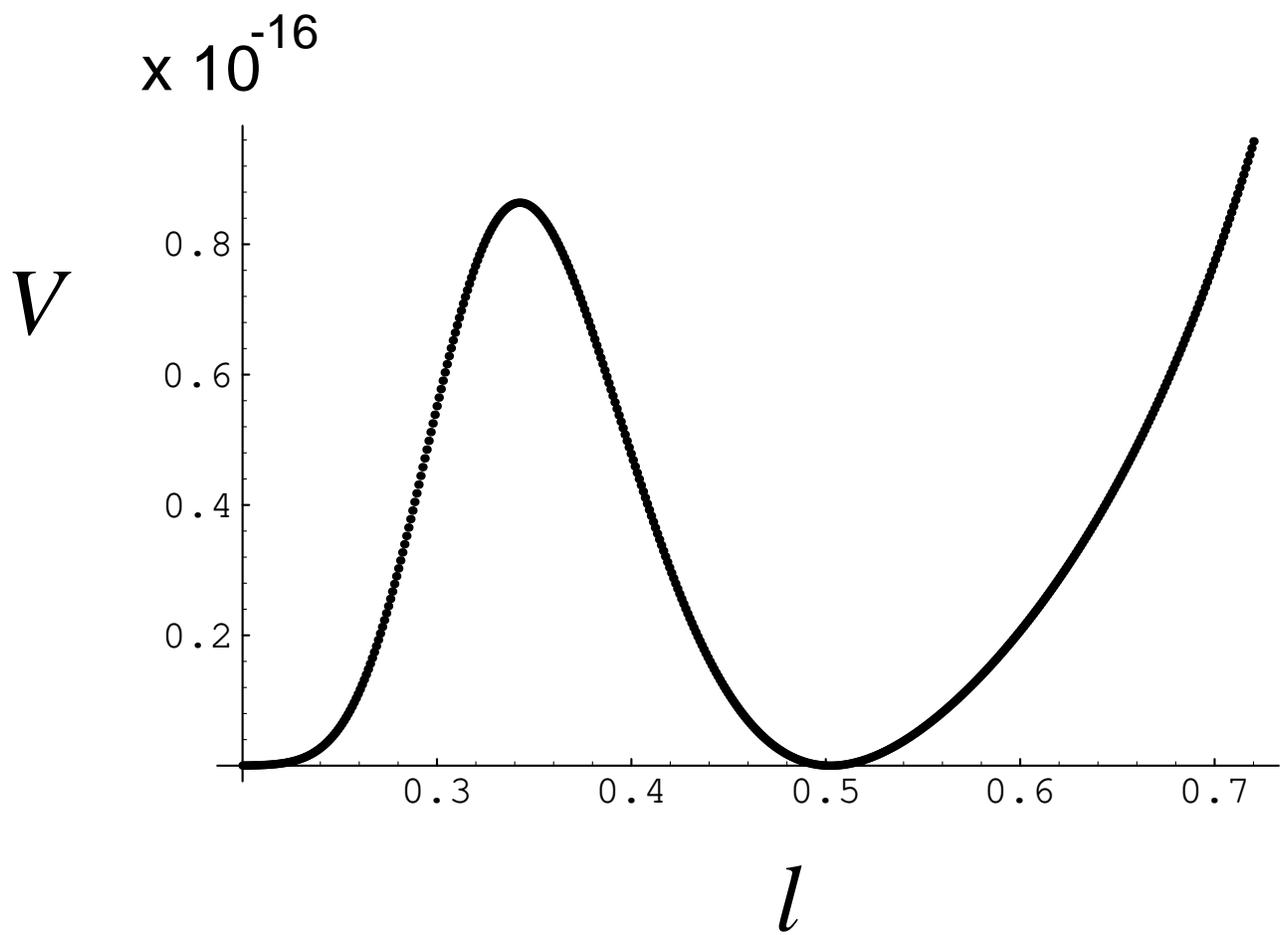
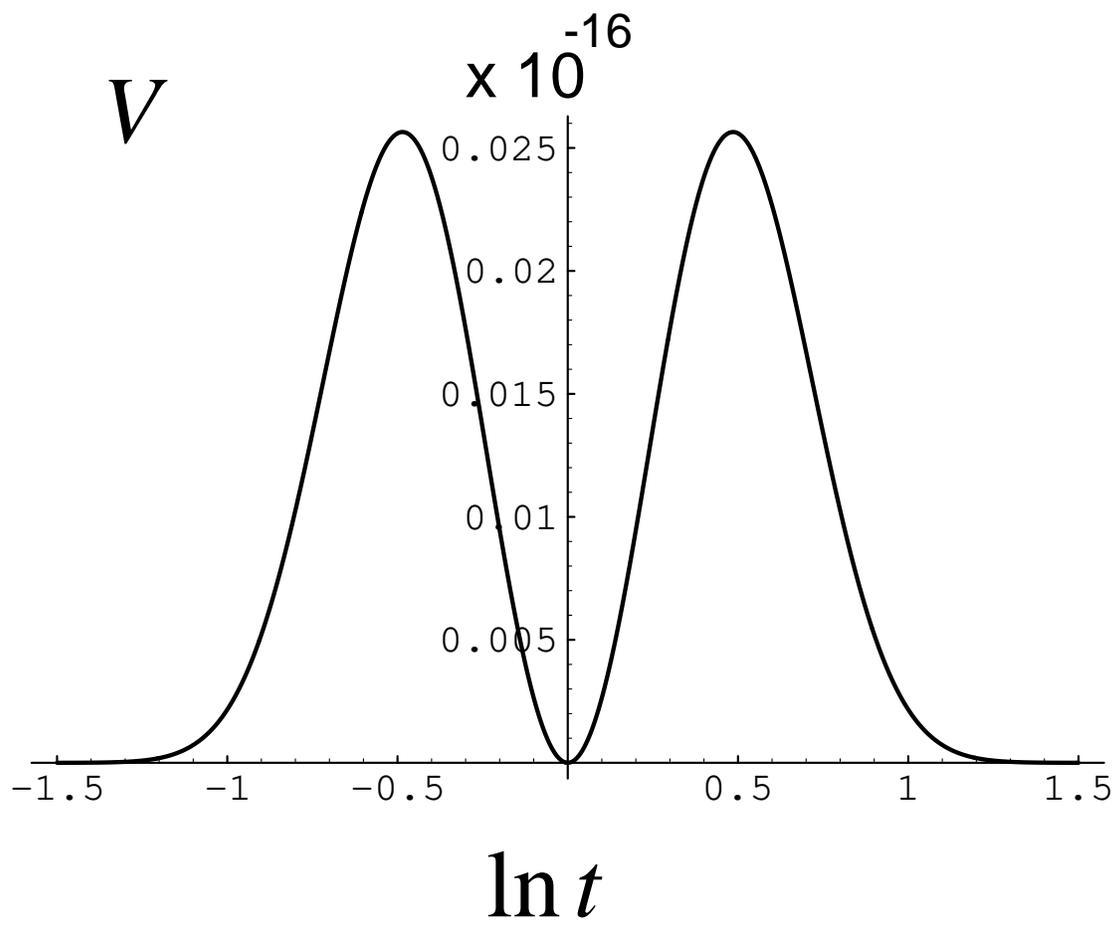
\end{document}